\documentstyle[12pt,epsfig]{article}

\input{epsf}

\newcommand{\be}{\begin{equation}}
\newcommand{\ee}{\end{equation}}
\newcommand{\bea}{\begin{eqnarray}}
\newcommand{\eea}{\end{eqnarray}}

\newcommand{\beq}{\begin{equation}}
\newcommand{\eeq}{\end{equation}}

\newcommand{\nn}{\nonumber}

\def\rhosq{\sqrt{1-\frac{4m^2}s}}
\def\srho{\sqrt{s(s-4m^2)}}

\def\fun#1#2{\lower3.6pt\vbox{\baselineskip0pt\lineskip.9pt
\ialign{$\mathsurround=0pt#1\hfil##\hfil$\crcr#2\crcr\sim\crcr}}}

\begin{document}

\title{Radiative decays of quarkonium states, momentum operator
expansion and nilpotent operators}

\author{A.V. Anisovich V.V. Anisovich, V.N. Markov,\\
 M.A. Matveev, V.A. Nikonov,
and A.V. Sarantsev}
\date{\today}
\maketitle

\begin{abstract}
We present the method of  calculation of the radiative decays of
composite quark--antiquark $Q\bar Q$  systems, with different
$J^{PC}$: $(Q\bar Q)_{in}\to \gamma(Q\bar Q)_{out}$.
The method is relativistic invariant, it is based on the  double
dispersion relation integrals over the masses of the composite mesons, it
can be used for the high spin particles and provides us with the
gauge invariant transition amplitudes. We apply this method to the case
when the photon is emitted by a constituent in the intermediate
state (additive quark model). We perform the momentum operator expansion
of the spin amplitudes for the decay processes. The problem of the
nilpotent spin operators is discussed.
\end{abstract}

\section{Introduction}

In this paper, we provide a consistent presentation of the method of
calculation of the radiative decays of the composite $Q\bar Q$ systems,
with arbitrary $J^{PC}$. We consider the radiative decays which are
realised through the transition $(Q\bar Q)_{in}\to \gamma(Q\bar Q)_{out}$
shown in Fig. 1 (additive quark model approach). The method is based on
the spectral integration over the masses of composite particles $(Q\bar
Q)_{in}$ and $(Q\bar Q)_{out}$ , it is relativistic and gauge invariant.
The obtained amplitude is determined by the quark wave functions of the
composite systems $(Q\bar Q)_{in}$ and $(Q\bar Q)_{out}$. The method,
in its substantial part, uses the spin operator expansion technique
developed previously in \cite{operator}.

\begin{figure}[t]
\centerline{\epsfig{file=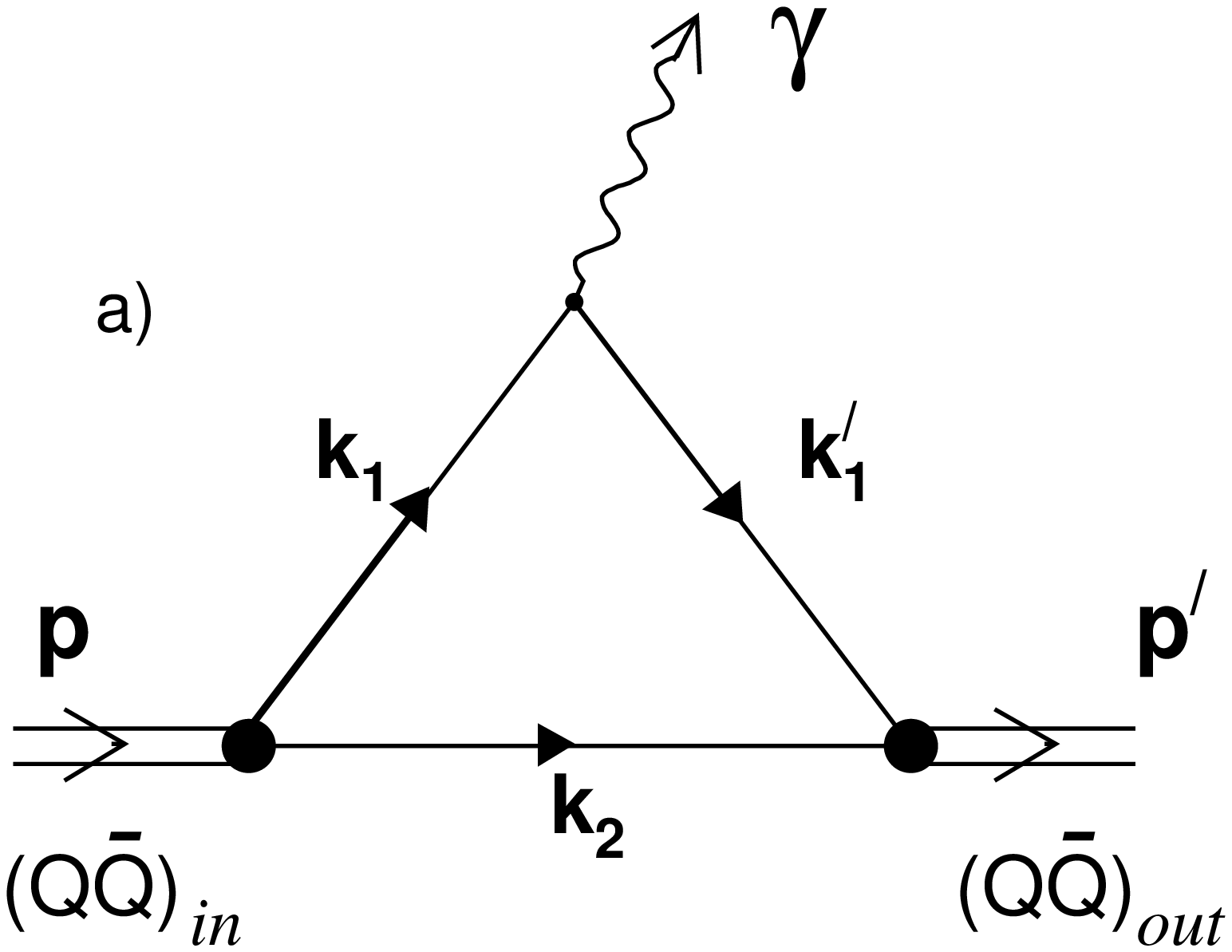,width=9cm}
            \epsfig{file=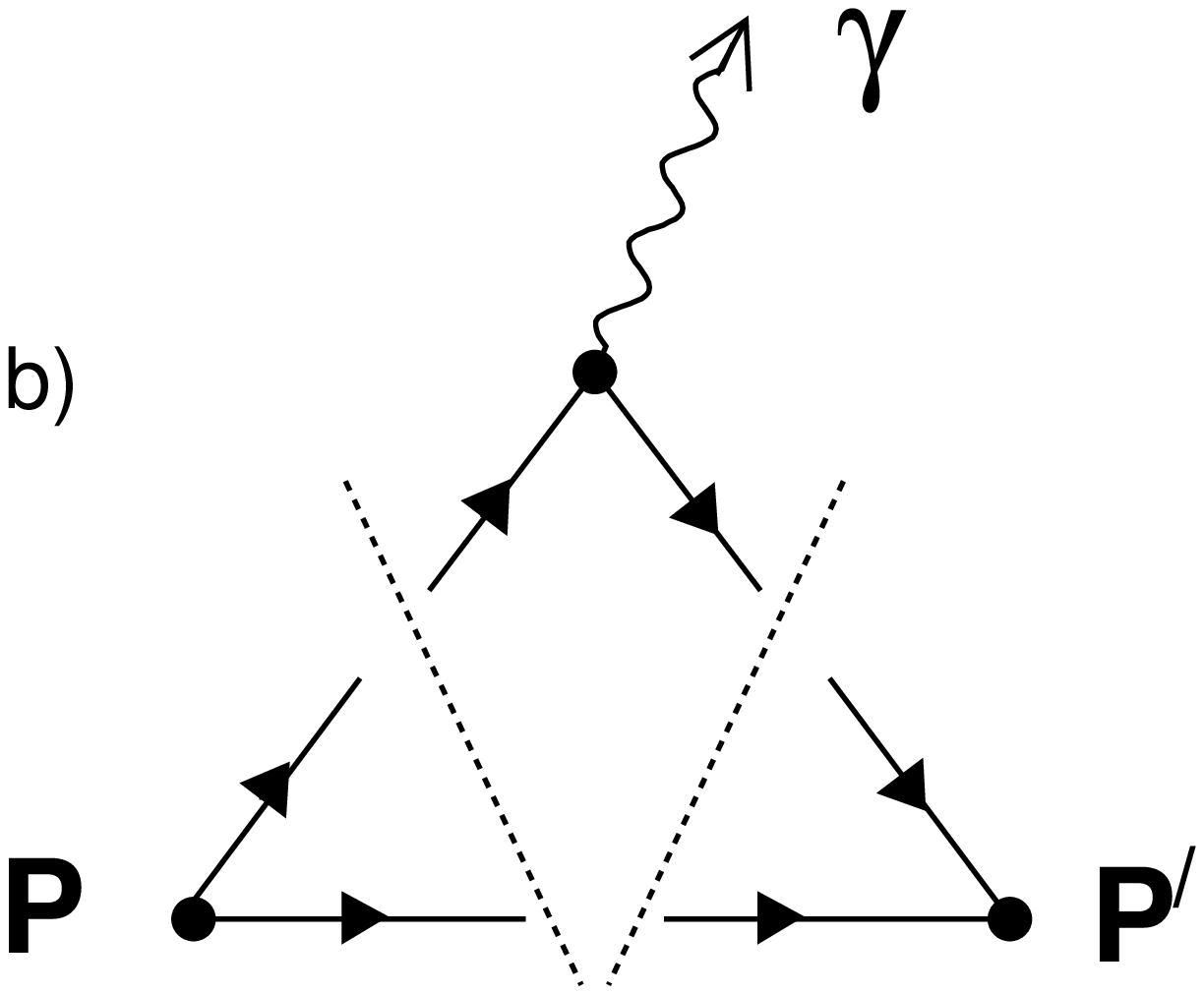,width=9cm}}
\vspace{0.5cm}
\caption{a) Triangle diagram for the radiative transition
$(Q\bar Q)_{in}\to \gamma(Q\bar Q)_{out}$
with $p^2=M^2_{in}$, $p'^2=M^2_{out}$ and $(p-p')^2=q^2$.
b) Cuttings of the triangle diagram for the double discontinuity of the
spectral integral with $P^2=s$, $P'^2=s'$ and $(P-P')^2=q^2$.}
\end{figure}

The consideration of triangle diagrams in terms of the spectral
integral over the mass of a composite particle, or  interacting
system, has a long history. Triangle diagrams appear at the
rescattering of the three-particle systems, and the energy dependences of
 corresponding amplitudes (on either total energy or one of pair
energies) were studied rather long ago, though in nonrelativistic
approximation, in the dispersion relation technique applied to the
analysis of the threshold singularities (see \cite{UFN-AA} and references
therein). The relativistic approximation was used for the extraction of
logarithmic singularity of the triangle diagram, e.g., see
\cite{Aitchison,AD} and references therein. Double dispersion relation
representation of the triangle diagram without accounting for
the spin structure was written in \cite{mos}.

In the consideration of radiative decays of the spin particles,
one of the most important point is a correct construction of the gauge
invariant spin operators that allows us to perform both the expansion
of the decay amplitude (written in terms of external variables) and
write down the double discontinuity of the spectral integral
(written in terms of the composite particle constituents). Such a procedure
had been realised for the deuteron in \cite{deut,deut-AS},
correspondingly, for the elastic scattering and photodisintegration
amplitude. A generalization of the method for the composite quark systems has
been performed in \cite{PR,EPJA,YF}.

There are two principal points which
 should be accounted for the processes shown in Fig. 1
considered in terms of the spectral integration technique:\\
(i) The amplitude of the process
$(Q\bar Q)_{in}\to \gamma(Q\bar Q)_{out}$  should be expanded in
a series in respect to a full set of spin operators, and this
expansion should be done in a uniform way  for both internal (quark)
and external boson states. The spin operators should be orthogonal,
and the
spectral integrals are to be written for the amplitudes related to these
orthogonal operators.\\
(ii) It  should be taken into account  that in the
processes with  the real photons (with photon four-momentum $q^2\to 0$)
the nilpotent spin operators appear, their norm being equal to zero
\cite{maxim}. Because of that, the representation of the
amplitudes may be different, that does not affect
the calculation result for partial widths.

The present paper was initiated by the study of the quarkonium
systems in terms of the spectral integral
Bethe--Salpeter equation --- the formulation of this equation for the
quark--antiquark systems as well as the discussion of its properties
may be found in \cite{BS}. Since the quark--antiquark interactions at
moderately large and large distances cannot be considered as
well-established, the treatment of quark--antiquark states should be
based on not only the knowledge on their masses but also on the wave
functions, and the source of  this latter  information is the
radiative decay processes. Therefore, we present the formulae for the
radiative transition amplitudes in the form convenient for their
simultaneous analysis  using the spectral integral Bethe--Salpeter
equation \cite{BS}.

The paper is organized as follows.

Section 2 is the introductory one. As an example, we use the
transitions $Q\bar Q(J^{PC}=0^{-+})\to \gamma+Q\bar Q(J^{PC}=1^{--})$
 and $Q\bar Q(J^{PC}=0^{++})\to \gamma+Q\bar
Q(J^{PC}=1^{--})$ studied before in
\cite{PR,EPJA,YF,maxim} and recall the method of
the amplitude representation in terms of the double spectral
integral. We also formulate the problem of
the momentum expansion of the spin amplitude in the case of nilpotent
operators.

In Section 3, the method is applied to the transition
 $Q\bar Q(2^{++})\to
\gamma+Q\bar Q(1^{--})$ and in Section 4 it is applied to
 $Q\bar Q(1^{++})\to \gamma+Q\bar
Q(1^{--})$.

In Conclusion, we emphasise that the cases
of  $Q\bar Q(2^{++})\to \gamma+Q\bar
Q(1^{--})$ and $Q\bar Q(1^{++})\to \gamma+Q\bar
Q(1^{--})$ are rather general and can be used  as a pattern
for the consideration of the
spectral integral representation of the amplitudes
$(Q\bar Q)_{in}\to \gamma+(Q\bar Q)_{out}$ for the $Q\bar Q$
states with arbitrary spin.

\section{Radiative transitions $P\to\gamma(q)V$ and $S\to\gamma(q)V$}

This Section is the introductory one: we consider here the meson
radiative transitions  $Q\bar Q(0^{-+})\to\gamma(q)+Q\bar Q( 1^{--})$
and $Q\bar Q(0^{++})\to\gamma(q)+Q\bar Q( 1^{--})$ (below the massive
mesons with $J^{PC}=0^{-+}, 1^{--},0^{++}$ are denoted as $P,V,S$,
correspondingly). These mesons were investigated previously
within the spectral integration technique \cite{EPJA,YF}. Using these
processes as examples, we demonstrate the basic principles of the
technique  used in subsequent sections for the more complicated
transitions. In parallel, we discuss the problem of the nilpotent
operators  that arise in the reactions with the photon emission.

\subsection{Transition $P\to\gamma(q)V$}

Here, we consider the transition $P\to
\gamma(q)V$ for the virtual photon. We write down the spin
operator for both initial mesons and quark intermediate states
in the triangle diagram, with the cuttings
shown in Fig. 1b. Then, we extract the invariant part of the
amplitude (form factor) and present it for the emission of real
photon expressed through the dispersion relation integral.

\subsubsection{Polarization vectors of the massive vector particle $V$
and photon}

The polarisations  of the vector meson, $\epsilon^{(V)}_{\beta}$, and
virtual photon, $\epsilon^{(\gamma(q))}_\alpha$,  are the transverse
vectors:
\bea
\epsilon^{(V)}_{\beta}p'_\beta\ =\ 0\ ,\qquad
\epsilon^{(\gamma(q))}_\alpha q_\alpha\ =\ 0\ .
\label{pv-1}
\eea
Here,  $q$ is the photon four-momentum and $p'$ is that of the
vector meson. The pseudoscalar meson momentum is denoted as
$p=q+p'$. The polarisation of the vector meson obeys
the completeness condition as follows:
\begin{eqnarray}
&& -\sum_{a=1,2,3}\epsilon^{(V)}_\alpha(a)
                  \epsilon^{(V){\bf +}}_\beta(a)\ =\
g^{\perp V}_{\alpha\beta}\ ,
\nonumber\\
&& g^{\perp V}_{\alpha\beta}\ =\
g_{\alpha\beta}- \frac{p'_\alpha p'_\beta}{p'^2}\ .
\label{pv-2}
\end{eqnarray}
Here,  $g^{\perp V}_{\alpha\beta}$ is the metric tensor operating in the
space orthogonal to the vector-meson momentum $p'$.

The polarisation vector of the real photon $(q^2=0)$ denoted as
$\epsilon^{(\gamma)}_\alpha$ has two independent components only,
they are orthogonal to the reaction
plane:
\beq \epsilon^{(\gamma)}_\alpha q_\alpha=0\ , \quad
\epsilon^{(\gamma)}_\alpha p_\alpha=0\ .
\label{apr3}
\eeq
Correspondingly, the completeness condition for the real photon reads:
\begin{eqnarray}
&& -\sum_{a=1,2}\epsilon^{(\gamma)}_\alpha(a)
\epsilon^{(\gamma){\bf +}}_{ \alpha'}(a)\ =\
g^{\perp\perp}_{\alpha\alpha'}\ , \nonumber\\ &&
g^{\perp\perp}_{\alpha\alpha'}\ =\ g_{\alpha\alpha'}- \frac{p_\alpha
p_{\alpha'}}{p^2}- \frac{q^{\perp}_{\alpha}q^{\perp}_{\alpha'}}{q^2_\perp}\ .
\label{apr4}
\end{eqnarray}
Here, $q^\perp$ is the orthogonal component of the photon momentum:
\begin{eqnarray}
&& q^{\perp}_{\alpha}=g^\perp_{\alpha\alpha'}
 q_{\alpha'}=q_\alpha-\frac{(pq)}{p^2}\,p_\alpha\ , \nonumber\\ &&
g^\perp_{\alpha\alpha'}=g_{\alpha\alpha'} -\frac{p_\alpha
p_{\alpha'}}{p^2}\ .
\label{apr5}
\end{eqnarray}
For virtual photon, $(q^2\neq0)$, the completeness condition for
polarisation vectors is written in  three-dimensional space:
\bea
\label{apr6}
-\sum_{a=1,2,3}\epsilon^{(\gamma^*)}_\alpha(a)\,
\epsilon^{(\gamma^*){\bf +}}_{\alpha'}(a)=
g^{\perp\gamma^*}_{\alpha\alpha'}\ ,
\\ \nonumber
g^{\perp\gamma^*}_{\alpha\alpha'}=
 g_{\alpha\alpha'}-\frac{q_\alpha q_{\alpha'}}{q^2}\, .
\eea

\subsubsection{Angular momenta of outgoing particles}

The angular momentum of outgoing particles depends on the relative momenta
of particles in their centre-of-mass system:
\beq \label{apr7}
g^\perp_{\alpha\alpha'}\cdot\frac12(q-p')_{\alpha'}=g^\perp_{\alpha\alpha'}
\left(q-\frac12\,p\right)_{\alpha'}=\ q^{\perp}_{\alpha}\ ,
\eeq
Following  \cite{operator}, we determine  the angular momenta  with the
help of operators $X^{(L)}_{\mu_1\cdots\mu_L}(q^\perp)$. Below, we give
an explicit form of these operators for the lowest states considered
here:
\bea \label{apr8}
 L\ =\ 0:&& X^{(0)}(q^\perp)\ =\ 1\ , \\
\nonumber L\ =\ 1:&& X^{(1)}_\mu(q^\perp)\ =\ q^{\perp}_{\mu}\ , \\
\nonumber L\ =\ 2:&& X^{(2)}_{\mu_1\mu_2}(q^\perp)\ =\
\frac32\left(q^{\perp}_{\mu_1}q^{\perp}_{\mu_2}
-\frac13g^\perp_{\mu_1\mu_2}q^2_\perp\right) \ .
\eea

\subsubsection{Amplitude for the decay $P\to\gamma V$}

The decay amplitude $P\to\gamma V$ (for the sake of definiteness, the real photon
is considered) is written as a product of the spin structure and form
factor:
\bea A_{P\to\gamma V}\ =\ \epsilon^{(\gamma)}_\alpha
\epsilon^{(V)}_\beta A^{(P\to\gamma V)}_{\alpha\beta}\ ,
\label{PgV-1}
\eea
where
\bea
A^{(P\to\gamma V)}_{\alpha\beta}\ =\ e\,
\varepsilon_{\alpha\beta\mu\nu} q^{\perp}_{\mu} p_\nu
F_{P\to\gamma V}(0) \ .
\label{PgV-2}
\eea
In (\ref{PgV-2}), the electron charge is singled out, and
$\varepsilon_{\alpha\beta\mu\nu}$ is the wholly antisymmetric tensor.
This expression works for the virtual photon transition
($\gamma\to\gamma^*$) with corresponding substitution:
$F_{P\to\gamma V}(0)\to F_{P\to\gamma V}(q^2)$.

In what follows the important role is played by the spin operator which
enters (\ref{PgV-2}):
$\varepsilon_{\alpha\beta\mu\nu}q^{\perp}_{\mu}p_\nu$
(or, in an abridged form,
$\varepsilon_{\alpha\beta q^\perp p}$). Since
$\varepsilon_{\alpha\beta p p}=0$, the spin operator can be
represented as follows:
\bea
S^{(P\to\gamma V)}_{\alpha\beta}(p,q)\ =\
\varepsilon_{\alpha\beta qp}\ .
\label{PgV-3}
\eea
Let us stress once more that this spin operator is valid for the
reaction with both real and virtual photons.

\subsubsection{Partial widths for $P\to\gamma V$ and $V\to\gamma P$}

The partial width for the decay $P\to\gamma V$ is determined as follows:
\bea
\label{PgV-4}
M_P \Gamma_{P\to\gamma V}&=& \int d\Phi_2(p;q,p')|
\sum_{\alpha\beta}A^{(P\to\gamma V)}_{\alpha\beta}|^2 \ =\
\frac{\alpha}{8}\frac{M_P^2-M_V^2}{M_P^2}\ |F_{P\to\gamma V}(0)|^2\ ,
\\ \nn
d\Phi_2(p;q,p')&=&\frac 12 \frac{d^3q}{(2\pi)^3\, 2q_{0}}
\frac{d^3p'}{(2\pi)^3\, 2p'_{0}} (2\pi)^{4}\delta^{(4)}(p-q -p')\ ,
\eea
where $M_V$ and $M_P$ are  the meson masses. The summation is
carried out over the photon and vector meson polarisations; in the
final expression $\alpha=e^2/4\pi=1/137$.

The same form factor gives us the partial width for the decay $V\to\gamma
P$:
\bea
\label{PgV-4a}
M_V \Gamma_{V\to\gamma P}&=&  \frac13\int d\Phi_2(p;q,p')|
\sum_{\alpha\beta}A^{(V\to\gamma P)}_{\alpha\beta}|^2 \ =\
\frac{\alpha}{24}\frac{M_V^2-M_P^2}{M_V^2}\ |F_{P\to\gamma V}(0)|^2\ .
\eea

\subsubsection{Double spectral integral representation of the triangle
diagram}

To derive double spectral integral for the form factor
$F_{P\to\gamma V}(0)$, one needs to calculate the double discontinuity
of the triangle diagram of Fig. 1b, where the cuttings are shown by
dotted lines. In the dispersion representation, the invariant energy in
the intermediate state differs from those of the initial and final
states. Because of that, in the double discontinuity $P\ne p$ and
$P'\ne p'$. The
following requirements are imposed on the momenta in the diagram of
Fig. 1b \cite{deut,PR}:
\bea (k_1+k_2)^2\ =\ P^2>4m^2\ ,\qquad
(k'_1+k_2)^2\ =\ P'^2>4m^2 \label{dsi-1} \eea at fixed $q^2$: \bea
(P'-P)^2\ =\ (k'_1-k_1)^2\ =\ q^2\ .
 \label{dsi-2}
\eea
Furthermore, in the spirit of the dispersion relation representation,
we denote $P^2=s$, $P'^2=s'$.

When we begin with Feynman diagram, the propagators should be
substituted by the residues in the poles, that is equivalent to
the replacement as follows:
$(m^2-k^2_i)^{-1}\to\delta(m^2-k^2)$.
Then, the double discontinuity of the amplitude
$A^{(P\to\gamma V)}_{\alpha\beta}$ becomes proportional to the three
factors:
\bea
\label{dsi-3}
disc_sdisc_{s'}A^{(P\to\gamma V(L))}_{\alpha\beta}\sim && Z_{P\to\gamma V}
G_P(s)G_{V(L)}(s')\times
\\ \nn
&&\times
d\Phi_2(P;k_1,k_2)d\Phi_2(P';k'_1,k'_2)
(2\pi)^32k_{20}\delta^3(\vec k'_2-\vec k_2)\times
\\ \nn
&&\times
Sp\left[i\gamma_5(\hat k_1+m)\gamma_\alpha^{\perp\gamma*}(\hat
k'_1+m)\hat G_\beta^{(1,L,1)}(k') (m-\hat k_2) \right] \ .
\eea
The first factor in the right-hand side of (\ref{dsi-3}) includes the
vertices: the quark charge factor $Z_{P\to\gamma V}$ (for the
one-flavour states $Z_{P\to\gamma V}=e_Q $) as well as transition
vertices $P\to Q\bar Q$ and $V\to Q\bar Q$ which are denoted as
$G_P(s)$ and $G_{V(L)}(s')$ (transition $V\to Q\bar Q$ is characterised
by two angular momenta $L=0,2$).

The second factor includes the space volumes of the two-particle states:
$d\Phi_2(P;k_1,k_2)$ and $d\Phi_2(P';k'_1,k'_2)$
that correspond to two cuts in the diagram of Fig. 1b (the space volume
is determined in (\ref{PgV-4})). The factor
$(2\pi)^32k_{20}\delta^3(\vec k'_2-\vec k_2)$ takes into account
the fact that one quark line is cut twice.

The third factor in (\ref{dsi-3}) is the trace coming from the summation
over the quark spin states. Since the spin factor in the transition
$V\to Q\bar Q$ may be of  two types (with dominant $S$- or dominant
$D$-wave), we have the following variants for
$\hat G^{(S,L,J)}_\beta(k')$:
\bea
\label{dsi-5}
L\ =\ 0:&&\hat G^{(1,0,1)}_\beta(k')\ =\ \gamma^{\perp V}_\beta\ ,
\\ \nn
L\ =\ 2:&&\hat G^{(1,2,1)}_\beta(k')\ =\ \sqrt{2} \gamma_{\beta'}
X^{(2)}_{\beta'\beta}(k')\,\ .
\eea
Here, $k'=(k'_1-k_2)/2$ is the momentum of outgoing quarks:
$k'\perp P'=k'_1+k_2$.

The total vertex  $\hat G^{V}_\beta(k')$ of the vector state is the sum
of  vertices for the components with $L=0$ and $L=2$:
\be
\hat G^{V}_\beta(k')\ =\hat G^{(1,0,1)}_\beta(k') G_{V(0)}(s')+  \hat
G^{(1,2,1)}_\beta(k') G_{V(2)}(s')
\ee
Correspondingly, there are two traces for  two
different transitions: $P\to\gamma V(0)$ and $P\to\gamma V(2)$:
\bea
\label{dsi-6}
Sp^{(P\to\gamma V(0))}_{\alpha\beta} &=&
-Sp[\hat G^{(1,0,1)}_\beta(k')(\hat k'_1+m)\gamma^{\perp\gamma*}_\alpha
(\hat k_1+m)i\gamma_5(-\hat k_2+m)]\ ,
\\ \nn
Sp^{(P\to\gamma V(2))}_{\alpha\beta} &=&
-Sp[\hat G^{(1,2,1)}_\beta(k')(\hat k'_1+m)\gamma^{\perp\gamma*}_\alpha
(\hat k_1+m)i\gamma_5(-\hat k_2+m)]\ .
\eea
Recall that $\gamma^{\perp\gamma*}_\alpha =
\gamma_{\alpha'}g^{\perp\gamma*}_{\alpha\alpha'}$,
see Eq. (\ref{apr6}).

To calculate the invariant form factor $F_{P\to\gamma V(L)}(q^2)$, we
should extract from (\ref{dsi-6}) the spin factor analogous to
$S^{(P\to\gamma V)}_{\alpha\beta}(q,p)$ given by Eq. (\ref{PgV-3}).
The total form factor is the sum as follows:
$F_{P\to\gamma V}(q^2)=F_{P\to\gamma V(0)}(q^2)+F_{P\to\gamma V(2)}(q^2)$.

For the $Q\bar Q$ quark states, this operator reads:
\bea
S^{(0^{-+}\to\gamma 1^{--})}_{\alpha\beta}(\tilde q,P')\
=\ \varepsilon_{\alpha\beta\tilde q P'}\ ,
\label{dsi-7}
\eea
where $\tilde q=P'-P$, while  $P'=k'_1+k_2$ and $P=k_1+k_2$.
Therefore, we have:
\bea
Sp^{(P\to\gamma V(L))}_{\alpha\beta} &=&
S^{(0^{-+}\to\gamma 1^{--})}_{\alpha\beta}(\tilde q,P')
S_{P\to\gamma V(L)}(s,s',q^2)\ ,
\label{dsi-81}
\eea
where
\bea
S_{P\to\gamma V(L)}(s,s',q^2)&=&
\frac{\left (Sp^{(P\to\gamma V(L))}_{\alpha\beta}
S^{(0^{-+}\to\gamma 1^{--})}_{\alpha\beta}(\tilde q,P')\right )}
{\left (S^{(0^{-+}\to\gamma 1^{--})}_{\alpha'\beta'}(\tilde q,P')
 S^{(0^{-+}\to\gamma 1^{--})}_{\alpha'\beta'}(\tilde q,P')\right )} \ .
\label{dsi-91}
\eea
As a result, we obtain:
\bea
\label{dsi-10}
S_{P\to\gamma V(0)}(s,s',q^2)&=& 4m \ ,
\\ \nn
S_{P\to\gamma V(2)}(s,s',q^2)&=&
\frac{m}{\sqrt 2}\left[(2m^2+s)-\frac{6ss'q^2}{\lambda(s,s',q^2)}\right]\ ,
\eea
with
\bea
\lambda=(s-s')^2-2q^2(s+s')+q^4 .
\label{dsi-11}
\eea
The double discontinuity of the amplitude (\ref{dsi-3}) is equal to
\bea
disc_s disc_{s'} A^{(P\to\gamma V(L))}_{\alpha\beta}& = &
S^{(0^{-+}\to\gamma 1^{--})}_{\alpha\beta}(\tilde q,P')\,
disc_s disc_{s'} F_{P\to\gamma V(L)}(s,s',q^2)\ ,
\label{dsi-12}
\eea
where
\bea
\label{dsi-13}
disc_s disc_{s'}F_{P\to\gamma V(L)}= &&Z_{P\to\gamma V}
G_P(s)G_{V(L)}(s') d\Phi_2(P;k_1,k_2)d\Phi_2(P';k'_1,k'_2)\times
\\ \nn
&&\times (2\pi)^32k_{20}\delta^3(\vec k'_2-\vec k_2)
S_{P\to\gamma V(L)}(s,s',q^2) \ .
\eea
It defines the form factor
through the dispersion integral as follows:
\bea
F_{P\to\gamma V(L)}(q^2)& = & \int\limits^{\infty}_{4m^2}
\frac{ds}{\pi} \int\limits^{\infty}_{4m^2}\frac{ds'}{\pi}
\frac{disc_s disc_{s'} F_{P\to\gamma V(L)}(s,s',q^2)}
{(s-M^2_P)(s'-M^2_{V(L)})}
\label{dsi-14}
\eea
We have written the expression  for $F_{P\to\gamma V(L)}(q^2) $
without subtraction terms, assuming that  the convergence of
(\ref{dsi-14}) is guaranteed by the vertices $G_P(s)$ and
$G_{V(L)}(s')$, see Eq. (\ref{dsi-13}). Furthermore, we define the wave
functions of the $Q\bar Q$ systems:
\bea
\psi_P(s)\ =\
\frac{G_P(s)}{s-M_P^2}\, ,\qquad
\psi_{V(L)}(s)\ =\
\frac{G_{V(L)}(s')}{s'-M_{V(L)}^2} \ .
 \label{dsi-15}
\eea
After
integrating over the momenta, in accordance with (\ref{dsi-13}), one can
represent (\ref{dsi-14}) in the following form:
 \be
 F_{P\to\gamma
V(L)}(q^2)=Z_{P\to\gamma V(L)} \int \limits_{4m^2}^\infty
\frac{dsds'}{16\pi^2} \psi_{P}(s)\psi_{V(L)}(s')
\frac{\Theta(-ss'q^2-m^2\lambda(s,s',q^2))}
{\sqrt{\lambda(s,s',q^2)}} S_{P\to\gamma V(L)}(s,s',q^2)\ ,
\ee
where $\Theta(X)$ is the  step-function: $\Theta(X)=1$ at
$X\ge 0$ and $\Theta(X)=0$ at $X<0$.

To calculate the integral at small $q^2$,
we make the  substitution:
\be
\label{dsi-16a}
s=\Sigma+\frac 12 zQ,\quad
s'=\Sigma-\frac 12 zQ,\quad
q^2=-Q^2\, ,
\ee
thus representing the form factor  as follows:
$$
F_{P\to\gamma V(L)}(-Q^2\to 0)=Z_{P\to\gamma V(L)}
\int \limits_{4m^2}^\infty
\frac{d\Sigma }{\pi} \psi_P(\Sigma)\psi_{V(L)}(\Sigma)
\int \limits_{-b}^{+b}\frac{ dz}{\pi}\;
\frac {S_{P\to\gamma V(L)}(\Sigma,z,-Q^2)}
{16\sqrt{\Lambda(\Sigma ,z ,Q^2)}}\ ,    \nn
$$
\be
\label{dsi-16b}
b=\sqrt{\Sigma (\frac{\Sigma}{m^2} -4)}, \qquad
\Lambda(\Sigma ,z, Q^2)=(z^2+4\Sigma) Q^2 \ .
\ee
After integrating over $z$ and substituting $\Sigma\to s$, the
form factors for $L=0,2$ read:
\bea
\label{PgV31}
F_{P\to\gamma V(0)}(0)&=&Z_{P\to\gamma V(0)}m
\int \limits_{4m^2}^\infty \frac{ds}{4\pi^2}
\psi_{P}(s)\psi_{V(0)}(s)\ln{\frac{s+\srho}{s-\srho}} \ ,
\\ \nn
F_{P\to\gamma V(2)}(0)&=&Z_{P\to\gamma V(2)}m
\int \limits_{4m^2}^\infty \frac{ds}{4\pi^2}
\psi_{P}(s)\psi_{V(2)}(s)\times
\\ \nn
&&\times
\left[(2m^2+s)\ln\frac{\sqrt{s}+\sqrt{s-4m^2}}{\sqrt{s}-\sqrt{s-4m^2}}
-3\sqrt{s(s-4m^2)}\right] \ .
\eea
The form factors (\ref{PgV31}) are expressed through
$\psi_{P}(s)$ and $\psi_{V(L)}(s)$ by the integral over $s$: these
expressions are sufficiently convenient for the combined fitting to
the spectral integral Bethe--Salpeter equation \cite{BS} and radiative
decay partial widths.

The total form factor is the sum as follows:
\be
F_{P\to\gamma V}(0)\ =\ F_{P\to\gamma V(0)}(0)+F_{P\to\gamma V(2)}(0)
\ee

The considered decay $P\to\gamma V$ offers rather simple
case for the representation of the process of Fig. 1 in the form
of the spectral integral, while  the
process $S\to\gamma V$ gives us a more complicated example --- there we face
the problem of the nilpotent spin
operators.

\subsection{Decay of the scalar meson  $S\to\gamma V$}

In the decay of the $0^{++}$ meson, the final particles can be  in the
$S$- and   $D$-wave states and the corresponding spin factors read:
\begin{eqnarray}
 S-\mbox{wave }:&& \quad\quad\quad (\epsilon^{(\gamma)}\epsilon^{(V)})\ ,
\nonumber\\
 D-\mbox{wave }:&& \quad\quad\quad X^{(2)}_{\alpha\beta}(q^\perp)\,
\epsilon^{(\gamma)}_\alpha\epsilon^{(V)}_\beta\ .
\label{apr11}
\end{eqnarray}
For real photon $q^{\perp}_{\alpha}\epsilon^{(\gamma)}_\alpha=0$,
so, using $X^{(2)}_{\alpha\beta}(q^\perp)$
from Eq. (\ref{apr8}), we see
that only the $g^\perp_{\alpha\beta}$ term
works in the $D$-wave. As a result, the $D$-wave operator,
\beq
 X^{(2)}_{\alpha\beta}(q^\perp)\,
\epsilon^{(\gamma)}_\alpha\epsilon^{(V)}_\beta =
-\frac12\,q^2_\perp\epsilon^{(\gamma)}_\alpha g^\perp_{\alpha\beta}
\epsilon^{(V)}_\beta=\
-\frac12\,q^2_\perp(\epsilon^{(\gamma)}\epsilon^{(V)}) \ ,
\label{apr12}
\eeq
gives us the structure similar to that of the $S$-wave. Therefore,
the amplitude of the real photon emission is determined by a single
spin factor in the amplitude: $(\epsilon^{(\gamma)}\epsilon^{(V)})$.
This factor may be represented as follows:
\beq
\epsilon^{(\gamma)}_\alpha
g^{\perp\perp}_{\alpha\beta} \epsilon^{(V)}_\beta\ ,
\label{apr13}
\eeq
where the metric tensor $g^{\perp\perp}_{\alpha\beta}$ is given by
(\ref{apr4}). Thence, we have the following spin operator for the
amplitude $S\to\gamma V$, when the real photon is emitted:
\beq
\label{apr14}
S^{(S\to\gamma V)}_{\alpha\beta}(p,q)\ =\ g^{\perp\perp}_{\alpha\beta}\
.
\eeq
This operator takes into consideration both $S$- and $D$-waves
in the transition $S\to\gamma V$.

However, the representation of spin operator
for the radiative transition $S\to\gamma V$ is not unique: along with
(\ref{apr14}), one may use a number of other forms.

\subsubsection{Transition amplitude $S\to\gamma V$ and ambiguities in
the representation of spin operator }

Using the spin operator (\ref{apr14}), we represent the amplitude
$S\to\gamma V$ as follows:
\be   \label{apr15a}
A^{(S\to\gamma V)}_{\alpha\beta}\ =\ S^{(S\to\gamma V)}_{\alpha\beta}(p,q)
F_{S\to\gamma V}(0),
\ee
where
$F_{S\to\gamma V}(0)$ is the transition form factor. Below we
demonstrate that the representation of the decay amplitude
(\ref{apr15a}) is not unique.

At $q^2=0$, the spin operator (\ref{apr14}) may be written as follows:
\beq
\label{apr15}
g^{\perp\perp}_{\alpha\beta}\equiv
g^{\perp\perp}_{\alpha\beta}(0) =g_{\alpha\beta} +\frac{p'^2}{(p'q)^2}\,
q_\alpha q_\beta-\frac1{(p'q)}(p'_\alpha q_\beta+q_\alpha p'_\beta)\ .
\eeq
Keeping in mind a consistent consideration of the case $q^2\neq 0$, here
we change the notation
$g^{\perp\perp}_{\alpha\beta}\to g^{\perp\perp}_{\alpha\beta}(0)$.

But it is also possible to apply another spin operator
in (\ref{apr15a}), that is done rather often:
\beq
\label{apr16}
S^{(S\to\gamma V)}_{\alpha\beta}(p,q)  \longrightarrow
\widetilde S_{\alpha\beta}\ =\
g_{\alpha\beta}- \frac{p'_\alpha q_\beta}{(p'q)}\ ,
\eeq
Here, as above, the index $\alpha$ relates to the photon
and  $\beta$ to vector meson. At $q^2=0$, this operator obeys
the requirement of transversality:
\beq
\label{apr17}
q_\alpha\widetilde S_{\alpha\beta}=\ 0\ , \quad
p'_\beta\widetilde S_{\alpha\beta}=\ 0\ ,
\eeq
so it may be equally applied to the  $S\to\gamma V$ amplitude.

The ambiguity in a representation of the amplitude is due to the
existence of the nilpotent operator $L_{\alpha\beta}(0)$,
\beq
\label{apr18}
L_{\alpha\beta}(0)L_{\alpha\beta}(0)\ =\ 0\ ,
\eeq
which is orthogonal to $g^{\perp\perp}_{\alpha\beta}(0)$:
\beq
\label{apr20}
g^{\perp\perp}_{\alpha\beta}(0)L_{\alpha\beta}(0)=\ 0\ .
\eeq
The operator $L_{\alpha\beta}(0)$ obeys the requirement of
transversity,
\beq
\label{apr19}
q_\alpha L_{\alpha\beta}(0)=\ 0\ ,
\quad L_{\alpha\beta}(0)p'_\beta=\ 0
\eeq
This operator can be
easily calculated using Eqs. (\ref{apr18}) and (\ref{apr20}) --- it is equal
to \cite{maxim}
\beq
\label{apr21}
L_{\alpha\beta}(0)=\
\frac{p'^2}{(p'q)^2}\,q_\alpha q_\beta- \frac1{(p'q)}\,q_\alpha
p'_\beta\ .
\eeq
It is seen from (\ref{apr16}) and (\ref{apr21}) that
$$
g^{\perp\perp}_{\alpha\beta}(0)= \widetilde
S_{\alpha\beta}+L_{\alpha\beta}(0)\ .
$$
Generally speaking, one can construct the
spin operator using any linear combination of
$g^{\perp\perp}_{\alpha\beta}(0)$ and $L_{\alpha\beta}(0)$:
\beq
\label{apr22}
S^{(S\to\gamma V)}_{\alpha\beta}=\ g^{\perp\perp}_{\alpha\beta}(0)+
C(p^2,p'^2)L_{\alpha\beta}(0)\ .
\eeq
Any of these operators may be equally applied to the construction of the
transition amplitude $S\to\gamma V$ for the emission of the real photon.

\subsubsection{Transition  $S\to\gamma^*V$ in the case of virtual photon
$q^2\neq0$}

For the virtual photon, $q^2\neq0$, the amplitude spin structure is
described by the  $g^{\perp\perp}_{\alpha\beta}$ and
$L_{\alpha\beta}$ operators which should be generalised for this very
case. In terms of $p'$ and $q$, these operators read:
\begin{eqnarray}
\label{apr23}
&& g^{\perp\perp}_{\alpha\beta}(p',q)\ =\ g_{\alpha\beta} +
\frac{q^2}{(p'q)^2-p'^2q^2}\,p'_\alpha p'_\beta\ +
\nonumber\\
&& +\quad \frac{p'^2}{(p'q)^2-p'^2q^2}\,q_\alpha q_\beta
-\frac{(p'q)}{(p'q)^2-p'^2q^2}\,(q_\alpha p'_\beta+p'_\alpha q_\beta)\ ,
\end{eqnarray}
and
\begin{eqnarray}
&& L_{\alpha\beta}(p',q)=\ \frac{q^2}{(p'q)^2-p'^2q^2}\,p'_\alpha
p'_\beta +\frac{p'^2}{(p'q)^2-p'^2q^2}\,q_\alpha q_\beta\ -
\nonumber\\
\label{apr24}
&&-\quad \frac{(p'q)}{(p'q)^2-p'^2q^2}\,q_\alpha p'_\beta-
\frac{p'^2q^2}{[(p'q)^2-p'^2q^2](p'q)}\,p'_\alpha q_\beta\ .
\end{eqnarray}
As is easy to see, these operators obey gauge invariance
and are orthogonal to each other. At
$q^2\to0$, the operators $g^{\perp\perp}_{\alpha\beta}(p',q)$ and
$L_{\alpha\beta}(p',q)$ transform into the formulae
(\ref{apr15}) and (\ref{apr21}), accordingly.

The transition amplitude $S\to\gamma^*V$
is determined by two form factors, namely,
\beq
\label{apr25}
A^{(S\to\gamma^*V)}_{\alpha\beta}=\ g^{\perp\perp}_{\alpha\beta}(p',q)
F_{tansverse}(q^2)+
L_{\alpha\beta}(p',q)F_{logitudinal}(q^2)\ .
\eeq
The first term in (\ref{apr25}) corresponds to the process with the
transverse polarised final state particles ($\gamma^*$ and
$V$), while the second one corresponds to the polarisations
lying in the reaction plane.

The operators $g^{\perp\perp}_{\alpha\beta}(p'q)$ and
$L_{\alpha\beta}(p',q)$ are singular. To avoid false
kinematical singularities in the amplitude
$A^{(S\to\gamma^*V)}_{\alpha\beta}$,  the poles in
$g^{\perp\perp}_{\alpha\beta}(p'q)$, $L_{\alpha\beta}(p',q)$
should be cancelled by zeros of the amplitude.

Instead of Eq. (\ref{apr25}), one can use another representation
of the decay amplitude, for example, that is based on the terms given in
(\ref{apr11}). Then, we have
\begin{eqnarray}
&& \hspace*{-4cm} S-\mbox{wave operator : }\quad\quad
q^{\perp\gamma^*}_{\alpha\alpha'}\, g^{\perp V}_{\alpha'\beta}\ ,
\nonumber\\
\label{apr26}
&& \hspace*{-4cm} D-\mbox{wave operator : }\quad\quad
q^{\perp\gamma^*}_{\alpha\alpha'}\,X^{(2)}_{\alpha'\beta'}(q^\perp)
g^{\perp V}_{\beta'\beta}\ .
\end{eqnarray}
Here, $g^{\perp V}_{\alpha\beta}$
is determined by Eq. (\ref{apr3}) and
$g^{\perp\gamma^*}_{\alpha\alpha'}$ by (\ref{apr6}). Note
that the convolution
$g^{\perp\gamma^*}_{\alpha\alpha'}g^{\perp V}_{\alpha'\beta}$ does not
coincide with
$g^{\perp\perp}_{\alpha\beta}(p',q)$ given at $q^2\neq0$ by Eq.
(\ref{apr23}).
There is another, much more important difference between the operators
given by Eq. (\ref{apr26}) and those given by Eqs. (\ref{apr23})
and
(\ref{apr24}): the operators (\ref{apr26}) are not orthogonal to each
other, in contrast to $g^{\perp\perp}_{\alpha\beta}(p',q)$ and
$L_{\alpha\beta}(p',q)$.  Indeed, the convolution of operators
(\ref{apr26}) gives:
\beq \label{apr27a}
g^{\perp\gamma^*}_{\alpha\alpha'}g^{\perp V}_{\alpha'\beta}
g^{\perp\gamma^*}_{\alpha\alpha'}X^{(2)}_{\alpha'\beta'}(q^\perp)
g^{\perp V}_{\beta'\beta}=\ -\frac{q^4_\perp}{3q^2p'^2} (p^2+p'^2+q^2).
\eeq
Non-orthogonality of the operators (\ref{apr26}) is due to the fact that
the convolution has been performed with the help of the metric tensors
$g^{\perp\gamma^*}_{\alpha\alpha'}$ and $g^{\perp V}_{\beta\beta'}$.
Had we operated with the normal metric tensor, namely, had we substituted in
(\ref{apr27a}) $g^{\perp\gamma^*}_{\alpha\alpha'}\to g_{\alpha\alpha'}$
and $g^{\perp V}_{\beta\beta'}\to g_{\beta\beta'}$,
we would have the orthogonal  $S$- and
$D$-wave operators. The metric tensors
$g^{\perp\gamma^*}_{\alpha\alpha'}$ and $g^{\perp V}_{\beta\beta'}$ in
(\ref{apr26}) allow us to fulfill the gauge invariance
 --- in this way, just due to the gauge invariance, the
orthogonality in the  $S$- and $D$-wave operators (\ref{apr26})
is broken. But, in the spectral representation of  form factors of the
composite systems, the
orthogonal operators are needed to avoid the double counting. This is the
reason why furthermore we deal with the orthogonal operators represented by
formulae (\ref{apr23}) and
(\ref{apr24}).

\subsubsection{Analytical properties of the amplitude at $q^2=0$}

Let us turn to the  real photon and discuss
analytical properties of  the amplitudes, namely, the cancellation of
kinematical singularities. The $A^{(S\to\gamma V)}_{\alpha\beta}$
amplitude reads:
\begin{eqnarray}
A^{(S\to\gamma V)}_{\alpha\beta} &=& \left[g_{\alpha\beta}+
\frac{4m^2_V}{(m^2_S-m^2_V)^2}q_\alpha q_\beta-\frac2{m^2_S-m^2_V}
(p'_\alpha q_\beta + q_\alpha p'_\beta) \right]F_{transverse}(0)\ +
\nonumber\\
 &&+\quad \left[\frac{4m^2_V}{(m^2_S-m^2_V)^2}\,q_\alpha q_\beta
-\frac2{m^2_S-m^2_V}\,q_\alpha p'_\beta\right]F_{longitudinal}(0)\ .
\label{apr27}
\end{eqnarray}
Here, we have used $2(p'q)=m^2_S-m^2_V$. To make nonsingular the
term in front of $q_\alpha q_\beta$  at $m^2_S\to m^2_V$, it is necessary
that
\beq
\label{apr28}
[F_{transverse}(0)+F_{longitudinal}(0)]_{m^2_S\to m^2_V}\ \sim\
(m^2_S-m^2_V)^2\ .
\eeq
This requirement is sufficient for the cancellation of kinematical
singularity in front of $q_\alpha p'_\beta$. However, to remove
kinematical singularity in the term  $p'_\alpha q_\beta$, the following
condition for $F_{transverse}(0)$ should be fulfilled:
\beq
\label{apr29}
F_{transverse}(0)\ \sim\ (m^2_S-m^2_V)\quad \mbox{ at  }\
(m^2_S-m^2_V)\rightarrow0\ .
\eeq
The constraint (\ref{apr28}) is in fact the requirement imposed
on $F_{longitudinal}(0)$, but the $F_{longitudinal}(0)$ itself, as was
noted above,
does not participate in the definition of the decay partial width
of the process $S\to\gamma V$.

The second constraint given by Eq. (\ref{apr29}) for
$F_{transverse}(0)$ is the principal one for the decay physics --- in
quantum mechanics it is known as Siegert's theorem \cite{siegert}, see
also the discussion in \cite{byers,achasov,antika}.

\subsubsection{Spin operator decomposition of the quark states
in the triangle diagram}

Now, let us consider the quark triangle diagram of Fig. 1. As was said
above, in the
 $Q\bar Q$ systems there are two possibilities to construct vector
mesons with the angular momenta
 $L=0$ and $L=2$. For the transitions   $V\to Q\bar Q(L)$, we apply the
 vertices introduced in (\ref{dsi-5}):
$\hat G^{(1,0,1)}_\beta$ and $\hat G^{(1,2,1)}_\beta$.
For the transition $S\to Q\bar Q(L)$, we use, following
\cite{operator}, the spin operator $mI$, where $I$ is the unit matrix.
The traces for two processes with the different vector-meson wave functions
($L=0,2$) read:
\bea
\label{SgV-1}
Sp^{(S\to\gamma V(0))}_{\alpha\beta} &=&
-Sp[\hat G^{(1,0,1)}_\beta(\hat k'_1+m)\gamma^{\perp\gamma*}_\alpha
(\hat k_1+m)mI(-\hat k_2+m)]\ ,
\\ \nonumber
Sp^{(S\to\gamma V(2))}_{\alpha\beta} &=&
-Sp[\hat G^{(1,2,1)}_\beta(\hat k'_1+m)\gamma^{\perp\gamma*}_\alpha
(\hat k_1+m)mI(-\hat k_2+m)]\ .
\eea
To calculate  the invariant form factor for the transverse polarised
final state particles (we denote it in an abridged form as
$F_{S\to\gamma V(L)}(q^2)$), one
should extract from (\ref{SgV-1}) the corresponding spin factor.
For the quark  states, this operator reads:
\bea
\label{SgV-2}
S^{(0^{++}\to\gamma 1^{--})}_{\alpha\beta}(\tilde q,P')\ =\
g^{\perp\perp}_{\alpha\beta} (\tilde q,P') \ .
\eea
Recall that $P=k'_1+k_2$ and $\tilde q=P-P'=k_1-k'_1$ .
We have:
\bea
\label{SgV-3}
Sp^{(S\to\gamma V(L))}_{\alpha\beta} &=&
S^{(0^{++}\to\gamma 1^{--})}_{\alpha\beta}(\tilde q,P')
S_{S\to\gamma V(L)}(s,s',q^2)\ ,
\eea
where
\bea
\label{SgV-4}
S_{S\to\gamma V(L)}(s,s',q^2)&=&
\frac{\left (Sp^{(S\to\gamma V(L))}_{\alpha\beta}
S^{(0^{++}\to\gamma 1^{--})}_{\alpha\beta}(\tilde q,P')\right )}
{\left (S^{(0^{++}\to\gamma 1^{--})}_{\alpha'\beta'}(\tilde
q,P')
 S^{(0^{++}\to\gamma 1^{--})}_{\alpha'\beta'}(\tilde q,P')\right )}\ .
\label{dsi-9}
\eea
The spin factors $S_{S\to\gamma V(L)}(s,s',q^2)$ at $L=0,2$
are equal to
\bea
S_{S\to\gamma V(0)}(s,s',q^2)&=&
-2m[(s-s'+q^2+4m^2)-\frac{4s'q^4}{\lambda (s,s',q^2)}]\ ,
\\ \nonumber
S_{S\to\gamma V(2)}(s,s',q^2)
&=&-\frac{m}{2\sqrt 2}[4m^4-2m^2(3s+s'-q^2)+s(s-s'+q^2)+\\ \nn  &+&
\frac{2ss'q^2}{\lambda (s,s',q^2)}
(16m^2+3q^2-s-3s')]\ ,
\eea
with $\lambda(s,s',q^2)$ given by (\ref{dsi-11}).

The form factor of the considered process takes the form:
\be
F_{S\to\gamma V(L)}(q^2)=Z_{S\to\gamma V}
\int \limits_{4m^2}^\infty \frac{dsds'}{16\pi^2}
\psi_{S}(s)\psi_{V(L)}(s')
\frac{\theta(-ss'q^2-m^2\lambda(s,s',q^2))}
{\sqrt{\lambda(s,s',q^2)}} S_{S\to\gamma V(L)}(s,s',q^2)\ .
\ee
To calculate the integral at $q^2\to 0$, we make, in a complete
similarity with the calculations done in Eqs. (\ref{dsi-16a}) and
(\ref{dsi-16b}), the  following substitution: $q^2=-Q^2$, $ s=\Sigma+
zQ/2$, $ s'=\Sigma-zQ/2$ and represent the form factor at small $Q$ as
follows:
\be
F_{S\to\gamma V(L)}(-Q^2\to 0)=Z_{S\to\gamma V}
\int \limits_{4m^2}^\infty
\frac{d\Sigma }{\pi} \psi_S(\Sigma)\psi_{V(L)}(\Sigma)
\int \limits_{-b}^{+b}\frac{ dz}{\pi}\;
\frac{S_{S\to\gamma V(L)}(\Sigma,z,-Q^2)}{16\sqrt{\lambda
(\Sigma ,z ,Q^2)}}\ ,
\ee
where $b=\sqrt{\Sigma (\Sigma /m^2 -4)}$ and
$\lambda(\Sigma ,z, Q^2)=(z^2+4\Sigma) Q^2 $.
After the integration over $z$
and substitution $\Sigma \to s$, we have:
\bea
F_{S\to\gamma V(0)}(0)&=&Z_{S\to \gamma V}\, \frac{m}{4\pi}
\int\limits_{4m^2}^{\infty}\frac{ds}{\pi}
\psi_S(s)\psi_{V(0)}(s)\, I_{S\to \gamma V}(s) ,
\\ \nn
F_{S\to\gamma V(2)}(0)&=&Z_{S\to \gamma V}\, \frac{m}{2\pi}
\int\limits_{4m^2}^{\infty}
\frac{ds}{\pi} \psi_S(s)\psi_{V(2)}(s)\,
(-s+4m^2)I_{S\to \gamma V}(s) ,
\\ \nn
I_{S\to \gamma V}(s) & =&\sqrt{s(s-4m^2)}-2m^2
\ln\frac{\sqrt{s}+\sqrt{s-4m^2}}{\sqrt{s}-\sqrt{s-4m^2}}\, .
\eea
The total form factor is equal to
\be
F_{S\to\gamma V}(0)\ =\ F_{S\to\gamma V(0)}(0)+F_{S\to\gamma V(2)}(0)\ .
\ee

\subsubsection{Partial widths for the decay processes with the emission
of real photon}

Similarly to the form factor calculations performed above, the
 partial width of the  scalar meson decay $S\to \gamma V$
 reads:
\bea
\label{PgS-pw}
M_S \Gamma_{S\to\gamma V}&=&
\int d\Phi_2(p;q,p')|
\sum_{\alpha\beta}A^{(S\to\gamma V)}_{\alpha\beta}|^2 \ =\
\frac{\alpha}{2}\frac{M_S^2-M_V^2}{M_S^2}\
|F_{S\to\gamma V}(0)|^2.
\eea
Recall that in the final expression $\alpha=e^2/4\pi =1/137$.
Likewise, partial width of the vector meson decay $V\to \gamma S$
is equal to:
\bea
\label{22}
M_V \Gamma_{V\to \gamma S}&=&\frac{\alpha}{6}\frac{M_V^2-M_S^2}{M_V^2}
\left |F_{S\to \gamma V}(0) \right |^2.
\eea
Here, we do not specify the quark structure of the vector meson
omitting the index $L$.

\subsubsection{Normalization conditions for wave functions
of $Q\bar Q$ states}

It is convenient to write the normalization conditions
for  $P$, $S$ and  $V$ meson wave functions using the charge form factor
of a meson:
\bea
F_{charge}(0)\ =\ 1\ . \label{norm-1}
\eea
The amplitude of the charge factor is defined by the diagram of
Fig. 1a, with
$(Q\bar Q)_{in}=(Q\bar Q)_{out}$. For $P$ and $S$ mesons,
the amplitude takes the form:
\bea A_\alpha(q)\ =\
e(p+p')_\alpha F_{charge}(q^2)\ , \label{norm-2} \eea
while $F_{charge}(q^2)$ can be calculated in the same way as the
transition form factors considered above. For the meson  $V$,
we consider the amplitude averaged over the spins of the massive vector
particle. At  $q^2=0$, it takes a form:
\bea A^{(V)}_{\alpha;\mu\mu}(q\to 0)\ =\ 3e(p+p')_\alpha
F^{(V)}_{charge}(0)\ . \label{norm-3} \eea

The normalisation conditions based on the formula (\ref{norm-2}) for $P$
and $S$ mesons read:
\bea
\label{norm-4}
1&=&\int\limits_{4m^2}^{\infty}\frac{ds}{16\pi^2}\ \psi_P^2(s)\ 2s\
\sqrt{\frac{s-4m^2}{s}}\ ,
\\ \nn
1&=&\int\limits_{4m^2}^{\infty}\frac{ds}{16\pi^2}\ \psi_S^2(s)\ 2m^2
\left(s-4m^2\right)\sqrt{\frac{s-4m^2}{s}}\ .
\eea

For the vector mesons $V$, the normalisation condition reads as
follows:
 \bea
\label{norm-5}
1 &=& W_{00}[V]+W_{02}[V]+W_{22}[V], \\
W_{00}[V]&=&\frac13\int\limits_{4m^2}^{\infty}
\frac{ds}{16\pi^2}\ \psi_{V(0)}^2(s)\ 4\left(s+2m^2\right)
\sqrt{\frac{s-4m^2}{s}}\ ,
\\ \nn
W_{02}[V]&=&\frac{\sqrt 2}{3}\int\limits_{4m^2}^{\infty}
\frac{ds}{16\pi^2}\ \psi_{V(0)}(s)\psi_{V(2)}(s)\ (s-4m^2)^2  \,
\sqrt{\frac{s-4m^2}{s}}\ ,
\\ \nn
W_{22}[V]&=&\frac23\int\limits_{4m^2}^{\infty}
\frac{ds}{16\pi^2}\ \psi_{V(2)}^2(s)\ \frac{(8m^2+s)(s-4m^2)^2}{16}
\sqrt{\frac{s-4m^2}{s}}\,.
\eea

In \cite{EPJA,YF}, the calculations of charge form factors are
explained in a more detail.

\section{Transitions $2^{++}$-$meson\to\gamma V$ and
 $1^{++}$-$meson\to\gamma V$}

In this Section, using the decays of the $2^{++}$-meson (denoted as
$T$) and  $1^{++}$-meson (denoted as $A$), we perform the treatment
which can be easily generalised for any spin particle.

\subsection{Transition $T\to\gamma V$}

To operate with the $2^+$-meson, we use the polarisation tensor
$\epsilon_{\mu\nu}(a)$ with five components $a=1,\ldots,5$.
This polarisation tensor,
being symmetrical  and traceless, obeys the completeness condition:
\bea
&&\sum_{a=1,\ldots,5}\epsilon_{\mu\nu}(a)\epsilon^{\bf +}_{\mu'\nu'}(a)=
\frac12\left(g^\perp_{\mu\mu'}g^\perp_{\nu\nu'}+g^\perp_{\mu\nu'}
g^\perp_{\nu\mu'}-\frac23g^\perp_{\mu\nu}g^\perp_{\mu'\nu'}\right)
=O^{\mu'\nu'}_{\mu\nu},
\nonumber \\
&&\sum_{a=1,\ldots,5}\epsilon_{\mu\nu}(a)\epsilon^{\bf +}_{\mu\nu}(a)\
=\ 5\ .
\label{TgV-1}
\eea
Here, $O^{\mu\nu}_{\mu'\nu'}$
is a standard projection operator $O^{\mu\nu}_{\mu'\nu'}$ for the
angular momentum $L=2$ which obeys the requirements:
$O^{\mu\nu}_{\mu''\nu''}O^{\mu''\nu''}_{\mu'\nu'}=O^{\mu\nu}_{\mu'\nu'}$
and $O^{\mu\nu}_{\mu'\mu'}=0$, see
\cite{operator} for more detail.

In terms of the polarisation tensor $\varepsilon_{\mu\nu}$
and vectors $\epsilon^{(\gamma^*)}_\alpha$, $ \epsilon^{(V)}_\beta$, the
independent
spin structures for the transitions with  virtual photon
$(q^2\neq0)$ read:
\begin{eqnarray}
&(1)\quad S\mbox{-wave : }
& \hspace*{3cm}
\epsilon_{\mu\nu} \epsilon^{(\gamma^*)}_\mu\epsilon^{(V)}_\nu\ ,
\nonumber\\
&(2)\quad D\mbox{-wave : }
& \hspace*{3cm} \epsilon_{\mu\nu}
X^{(2)}_{\mu\nu}(q^\perp)(\epsilon^{(\gamma^*)}\epsilon^{(V)})\ ,
\nonumber\\
&(3)\quad D\mbox{-wave  :}
& \hspace*{3cm} \epsilon_{\mu\nu}X^{(2)}_{\nu\beta}(q^\perp)
\epsilon^{(\gamma^*)}_\mu\epsilon^{(V)}_\beta\ ,
\nonumber\\
&(4)\quad D\mbox{-wave  :}
& \hspace*{3cm} \epsilon_{\mu\nu}X^{(2)}_{\nu\alpha}(q^\perp)
\epsilon^{(\gamma^*)}_\alpha \epsilon^{(V)}_\mu\ ,
\nonumber\\
&(5)\quad G\mbox{-wave :}
& \hspace*{3cm}
\epsilon_{\mu\nu}X^{(4)}_{\mu\nu\alpha\beta}(q^\perp)
\epsilon^{(\gamma^*)}_\alpha \epsilon^{(V)}_\beta\ .
\label{TgV-2}
\end{eqnarray}
Correspondingly, we have five independent form factors which describe the
transition $2^{++}$-$meson$ $\to$ $\gamma^* V$. But, for the real
photon $(q^2=0)$, the number of independent form factors is reduced to
three and, keeping this fact in mind, we consider below the
production of the transverse polarised photon.

\subsubsection{Spin operators for transverse
polarised photon, $2^+$-$meson\to\gamma^{*\perp} V$ }

Here, we consider the transverse polarised photon, though with
$q^2\neq 0$. We introduce the following spin operators, which correspond
to the spin structures (\ref{TgV-2}):
\bea
\label{TgV-3}
S_{\mu\nu,\alpha\beta}^{(1)}&=& O^{\mu'\nu'}_{\mu\nu}
g^{\perp\perp}_{\mu'\alpha}g^{\perp V}_{\nu'\beta}\ ,
\\
S_{\mu\nu,\alpha\beta}^{(2)}&=&-\frac 1{q_\perp^2}
O^{\mu'\nu'}_{\mu\nu}X^{(2)}_{\mu'\nu'}
(q^\perp)g^{\perp\perp}_{\alpha\alpha'}g^{\perp V}_{\alpha'\beta}=
-\frac 1{q_\perp^2}X^{(2)}_{\mu\nu}(q^\perp)g^{\perp\perp}_{\alpha\beta}\ ,
\nn \\
S_{\mu\nu,\alpha\beta}^{(3)}&=& -\frac 1{q_\perp^2}O^{\mu'\nu'}_{\mu\nu}
X^{(2)}_{\nu'\beta'}(q^\perp)g^{\perp\perp}_{\mu'\alpha}
g^{\perp V}_{\beta'\beta}\ ,
\nn  \\
S_{\mu\nu,\alpha\beta}^{(4)}&=& -\frac 1{q_\perp^2}O^{\mu'\nu'}_{\mu\nu}
X^{(2)}_{\nu'\alpha'}(q^\perp) g^{\perp\perp}_{\alpha'\alpha}
g^{\perp V}_{\mu'\beta}\ ,
\nn  \\
S_{\mu\nu,\alpha\beta}^{(5)}&=& \frac 1{q_\perp^4}O^{\mu'\nu'}_{\mu\nu}
X^{(4)}_{\mu'\nu'\alpha'\beta'}(q^\perp)
g^{\perp\perp}_{\alpha'\alpha} g^{\perp V}_{\beta'\beta} \ . \nn
\eea
Recall that here
$q^{\perp}_{\alpha}=g^\perp_{\alpha\alpha'}
q_{\alpha'}=q_\alpha-p_\alpha(pq)/p^2$ and
$g^\perp_{\alpha\alpha'}=g_{\alpha\alpha'} -p_\alpha
p_{\alpha'}/p^2$.

Let us demonstrate the method of construction of these operators by
considering the $G$-wave spin structure from (\ref{TgV-2}),
$\epsilon_{\mu'\nu'}X^{(4)}_{\mu'\nu'\alpha'\beta'}(q^\perp)
\epsilon^{(\gamma^*)}_{\alpha'} \epsilon^{(V)}_{\beta'} $.
It should be multiplied by the polarisations
$\epsilon_{\mu\nu}^+(a)$,
$\epsilon^{(\gamma^*\perp +)}_\alpha (b)$, $\epsilon^{(V)+}_\beta(c)$,
with the summation over $a,b,c$:
\bea
\label{TgV-4}
\sum_{a,b,c}
\epsilon_{\mu\nu}^+(a)\epsilon_{\mu'\nu'}(a)
X^{(4)}_{\mu'\nu'\alpha'\beta'}(q^\perp)
\epsilon^{(\gamma^*\perp)}_{\alpha'}(b)
\epsilon^{(\gamma^*\perp)+}_\alpha(b)
 \epsilon^{(V)}_{\beta'}(c)\epsilon^{(V)+}_\beta(c),\nn
\eea
that gives us $S_{\mu\nu,\alpha\beta}^{(5)}$, because the convolution
$\epsilon^{(\gamma^*\perp)}_{\alpha'}(b)
\epsilon^{(\gamma^*\perp)+}_\alpha(b)$ results in
$g^{\perp\perp}_{\alpha'\alpha}$.

The operators (\ref{TgV-3}) should be orthogonalised as follows:
\bea
\label{TgV-6}
S^{(\perp 1)}_{\mu\nu,\alpha\beta}(p',q)&=&
S^{(1)}_{\mu\nu,\alpha\beta}\ ,
\\ \nn
S^{(\perp 2)}_{\mu\nu,\alpha\beta}(p',q)&=&
S^{(2)}_{\mu\nu,\alpha\beta}-S^{(\perp 1)}_{\mu\nu,\alpha\beta}(p',q)
\frac{\left(S^{(\perp 1)}_{\mu'\nu',\alpha'\beta'}(p',q)
            S^{(2)}_{\mu'\nu',\alpha'\beta'}\right)}
     {\left(S^{(\perp 1)}_{\mu'\nu',\alpha'\beta'}(p',q)
            S^{(\perp 1)}_{\mu'\nu',\alpha'\beta'}(p',q)\right)} \ ,
\\ \nn
S^{(\perp 3)}_{\mu\nu,\alpha\beta}(p',q)&=&
S^{(3)}_{\mu\nu,\alpha\beta}-S^{(\perp 1)}_{\mu\nu,\alpha\beta}(p',q)
\frac{\left(S^{(\perp 1)}_{\mu'\nu',\alpha'\beta'}(p',q)
            S^{(3)}_{\mu'\nu',\alpha'\beta'}\right)}
     {\left(S^{(\perp 1)}_{\mu'\nu',\alpha'\beta'}(p',q)
            S^{(\perp 1)}_{\mu'\nu',\alpha'\beta'}(p',q)\right)}
\\ \nn
&-&S^{(\perp 2)}_{\mu\nu,\alpha\beta}(p',q)
\frac{\left(S^{(\perp 2)}_{\mu'\nu',\alpha'\beta'}(p',q)
            S^{(3)}_{\mu'\nu',\alpha'\beta'}\right)}
     {\left(S^{(\perp 2)}_{\mu'\nu',\alpha'\beta'}(p',q)
            S^{(\perp 2)}_{\mu'\nu',\alpha'\beta'}(p',q)\right)} \ .
\eea
In this way, we construct three operators, $i=1,2,3$.
The operators $S^{(\perp 4)}_{\mu\nu,\alpha\beta}$ and
$S^{(\perp 5)}_{\mu\nu,\alpha\beta}$ are nilpotent at
$q^2=0$, so we do not present explicit expressions for them here but
concentrate our attention on the calculation of the amplitude for the
emission of the real photon.

The results of calculation of the spin operator convolutions are
presented in Appendix 2. Here, we give the orthogonalised operator
norms, which determine the decay partial width:
\bea
S_{\mu\nu,\alpha\beta}^{(\perp a)}(p',q)
S_{\mu\nu,\alpha\beta}^{(\perp b)}(p',q)&\equiv&
z^\perp_{ab}(M^2_T,M^2_V,q^2), \\
z^\perp_{11}(M^2_T,M^2_V,0)&=&\frac
{3 M^4_T+34  M^2_T M^2_V+3 M^4_V}{12
M^2_T M^2_V}\ , \nn \\
 z^\perp_{22}(M^2_T,M^2_V,0)&=&9\frac{M^4_T+10
M^2_T M^2_V+M^4_V} {3 M^4_T+34  M^2_T
M^2_V+3 M^4_V}\ ,
 \nn \\
z^\perp_{33}(M^2_T,M^2_V,0)&=&\frac 92\, \frac
{(M^2_T+M^2_V)^2}{M^4_T+10 M^2_T M^2_V+M^4_V}\ .
 \nn
\eea

\subsubsection{Calculation of the transition amplitude $T(L)\to\gamma
V(L')$ for the emission of the real photon}

The transition amplitude of the $T\to \gamma V$ decay can be written
using the operators (\ref{TgV-6}) as follows:
\be
\label{TgV-7}
A^{(T(L)\to\gamma V(L'))}_{\mu\nu;\alpha\beta}\ =\
\sum_{i=1,2,3}
S^{(\perp i)}_{\mu\nu;\alpha\beta}(p',q)
F^{(i)}_{T(L)\to\gamma V(L')}(0) \ ,
\ee
where  $F^i_{T(L)\to\gamma V(L')}(0)$ are the form factors at $q^2=0$.

Considering the double
discontinuity  related to  Fig. 1b, we should expand the following
traces in a series over the spin operators:
\bea \label{TgV-8}
Sp_{\mu\nu,\alpha\beta}^{(T(1)\to\gamma V(0))}& =&
-Sp\left[\hat G^{(1,0,1)}_\beta(\hat{k'_1}+m)\gamma_\alpha^{\perp\gamma}
(\hat{k_1}+m)T^{(1)}_{\mu\nu}(k)(-\hat{k_2}+m)\right],
\\ \nn
Sp_{\mu\nu,\alpha\beta}^{(T(1)\to\gamma V(2))}& =&
-Sp\left[\hat G^{(1,2,1)}_\beta(\hat{k'_1}+m)\gamma_\alpha^{\perp\gamma}
(\hat{k_1}+m)T^{(1)}_{\mu\nu}(k)(-\hat{k_2}+m)\right],
\\ \nn
Sp_{\mu\nu,\alpha\beta}^{(T(3)\to\gamma V(0))}& =&
-Sp\left[\hat G^{(1,0,1)}_\beta(\hat{k'_1}+m)\gamma_\alpha^{\perp\gamma}
(\hat{k_1}+m)T^{(3)}_{\mu\nu}(k)(-\hat{k_2}+m)\right],
\\ \nn
Sp_{\mu\nu,\alpha\beta}^{(T(3)\to\gamma V(2))}& =&
-Sp\left[\hat G^{(1,2,1)}_\beta(\hat{k'_1}+m)\gamma_\alpha^{\perp\gamma}
(\hat{k_1}+m)T^{(3)}_{\mu\nu}(k)(-\hat{k_2}+m)\right]\ .
\eea
Recall that here we have used the notations
$ \gamma_\alpha^{\perp\gamma}\ =\ g_{\alpha\alpha'}^{\perp\perp}\gamma_{\alpha'}$,
$ k\ =\ (k_1-k_2)/2$, $ k'\ =\ (k'_1-k_2)/2$
and vertices
$\hat G^{(1,0,1)}_\beta$, $\hat G^{(1,2,1)}_\beta$ given in
Eq. (\ref{dsi-5}).

The operators for the transitions $2^{++}$-$meson \to Q\bar Q(L)$
for $L=1,3$ read:
\bea
T^{(1)}_{\mu\nu}(k)&=&\frac{3}{\sqrt 2}\left[k_\mu\gamma_\nu
+k_\nu\gamma_\mu-\frac 23 g_{\mu\nu}^\perp \hat{k}\right],
\\ \nn
T^{(3)}_{\mu\nu}(k)&=&\frac{5}{\sqrt 2} \left [k_{\mu}k_{\nu}\hat k
-\frac15 k^2(g^\perp_{\mu\nu}\hat k +\gamma_{\mu}k_{\nu}
+k_{\mu}\gamma_{\nu})\right ] .
\eea
To calculate invariant form factors $F^{(i)}_{T\to\gamma V(L)}(q^2)$, we
should expand  the traces of (\ref{TgV-8}) in a series over the spin
operators for the
quark states $S^{(\perp i)}_{\mu\nu,\alpha\beta}(P',\tilde q)$  and
extract the invariant factors:
\bea
\label{dsi-8}
Sp^{(T(L)\to\gamma V(L'))}_{\mu\nu,\alpha\beta} &=& \sum_{i=1,2,3}
S^{(\perp i)}_{\mu\nu,\alpha\beta}(P',\tilde q)
S^{(i)}_{T(L)\to\gamma V(L')}(s,s',q^2)\ ,
\eea
Let us emphasise that in (\ref{dsi-8}) the spin operators depend on the
intermediate-state quark variables, $P'$ and $\tilde q$.

The invariant spin factors read:
\bea
\label{dsi-9}
S^{(\perp i)}_{T(L)\to\gamma V(L')}(s,s',q^2)&=&
\frac{\left (Sp^{(T(L)\to\gamma V(L'))}_{\mu\nu,\alpha\beta}
S^{(\perp i)}_{\mu\nu,\alpha\beta}(P',\tilde q)\right )}
{\left (S^{(\perp i)}_{\mu\nu,\alpha\beta}(P',\tilde q)
 S^{(\perp i)}_{\mu\nu,\alpha\beta}(P',\tilde q)\right )}\ ,
\eea
where $i=1,2,3$.  Invariant spin factors determine the
form factors in a standard way:
\bea
F^{(i)}_{T(L)\to\gamma V(L')}(q^2)=&&Z_{T\to\gamma V}
\int \limits_{4m^2}^\infty \frac{dsds'}{16\pi^2}
\psi_{T(L)}(s)\psi_{V(L')}(s')\times
\\ \nn
&&\times\,\frac{\theta(-ss'q^2-m^2\lambda(s,s',q^2))}
{\sqrt{\lambda(s,s',q^2)}} S^{(\perp i)}_{T(L)\to\gamma V(L')}(s,s',q^2)
\eea
To calculate the integral at $q^2\to 0$, we make, as before (see
Eqs. (\ref{dsi-16a}) and (\ref{dsi-16b})), the  following substitution:
$q^2=-Q^2$, $ s=\Sigma+ zQ/2$, $ s'=\Sigma-zQ/2$ and perform the
integration over $z$. We have:
\bea
F_{T(L)\to\gamma V(L')}^{(i)}&=&Z_{T(L)\to\gamma V(L')}
\int\limits_{4m^2}^\infty \frac{ds}{16\pi^2}
\psi_{T(L)}(s)\psi_{V(L')}(s) J^{(i)}_{T(L)\to\gamma V(L')}(s)  \, .
\eea
Here,
\bea
S^{(1)}_{T(1)\to\gamma V(0)}(s)&=&
-\frac{\sqrt 3}{5} (8m^2+3s) I^{(1)}_{T\to\gamma V}(s)\ ,
\nn \\
S^{(2)}_{T(1)\to\gamma V(0)}(s)&=&
\frac23S^{(3)}_{T(1)\to\gamma V(0)}(s)
=-\frac{2}{3\sqrt{3}} I^{(2)}_{T\to\gamma V}(s)\ ,
\nn \\
S^{(1)}_{T(1)\to\gamma V(2)}(s)&=&
-\frac{\sqrt{6}}{40} (16m^2-3s)(4m^2-s) I^{(1)}_{T\to\gamma V}(s)\ ,
\nn \\
S^{(2)}_{T(1)\to\gamma V(2)}(s)&=&
\frac23S^{(3)}_{T(1)\to\gamma V(2)}(s)
= -\frac{\sqrt{2}}{12\sqrt{3}}(8m^2+s) I^{(2)}_{T\to\gamma V}(s)\ ,
\nn \\
S^{(1)}_{T(3)\to\gamma V(0)}(s)&=&
-\frac{3\sqrt{2}}{20} (4m^2-s)^2 I^{(1)}_{T\to\gamma V}(s)\ ,
\nn \\
S^{(2)}_{T(3)\to\gamma V(0)}(s)&=&
\frac23S^{(3)}_{T(3)\to\gamma V(0)}(s)
= -\frac{\sqrt{2}}{18}(6m^2+s) I^{(2)}_{T\to\gamma V}(s)\ ,
\nn \\
S^{(1)}_{T(3)\to\gamma V(2)}(s)&=&
-\frac{3}{80} (4m^2-s)^2(8m^2+s) I^{(1)}_{T\to\gamma V}(s)\ ,
\nn \\
S^{(2)}_{T(3)\to\gamma V(2)}(s)&=&
\frac 23S^{(3)}_{T(3)\to\gamma V(2)}(s)
= -\frac{1}{72} (16m^2-3s)(4m^2-s) I^{(2)}_{T\to\gamma V}(s)\ ,
\eea
where
\bea
\label{31b}
I^{(1)}_{T\to\gamma V}(s)&=&
2m^2\ln\frac{\sqrt{ s}+\sqrt {s-4m^2}}
            {\sqrt{ s}-\sqrt {s-4m^2}}-\sqrt {s(s-4m^2)},
\\ \nn
I^{(2)}_{T\to\gamma V}(s)&=&
m^2(m^2+s)\ln\frac{\sqrt{ s}+\sqrt {s-4m^2}}{\sqrt{ s}-\sqrt {s-4m^2}}
-\frac 1{12}\sqrt {s(s-4m^2)}(s+26m^2)\ .
\eea
Total form factor is a sum over four terms:
\be
F^{(i)}_{T\to\gamma V}=\sum\limits_{L,L'} F_{T(L)\to\gamma V(L')}^{(i)} .
\label{31a}
\ee

\subsubsection{Normalisation conditions and partial widths}

 The normalisation condition for tensor mesons reads:
\bea
1&=&W_{11}[T]+W_{13}[T]+W_{33}[T],\\
W_{11}[T]&=& \frac15\int \limits_{4m^2}^\infty \frac {ds}{16\pi^2}
\; \psi_{T(1)}^2(s)\; \frac 12 (8m^2+3s)(s-4m^2)\rhosq ,
\\ \nn
W_{13}[T]&=& \frac15\int \limits_{4m^2}^\infty \frac {ds}{16\pi^2} \;
\psi_{T(1)}(s)\psi_{T(3)}(s)\;
\frac {\sqrt 3}{2\sqrt 2} (s-4m^2)^3\, \rhosq ,
\\ \nn
W_{33}[T]&=& \frac15\int \limits_{4m^2}^\infty \frac {ds}{16\pi^2} \;
\psi_{T(3)}^2(s)\; \frac 1{16} (6m^2+s)(s-4m^2)^3\, \rhosq \ .
\eea
Partial width of the decay $T\to\gamma V$ is equal to:
\bea
m_T \Gamma_{T\to\gamma V}&= &
e^2\int d\Phi_2(p;,q,p') \frac 15 \sum \limits_{\mu\nu,\alpha\beta}
|A_{\mu\nu,\alpha\beta}|^2
\\ \nn
&=&\frac{\alpha}{20} \frac{m_T^2-m_V^2}{m_T^2}
\left[z^\perp_{11}(M^2_T,M^2_V,0)\ (F^{(1)}_{T\to\gamma V}(0))^2 +
\right .
\\ \nn
&& +\left .
z^\perp_{22}(M^2_T,M^2_V,0)\ (F^{(2)}_{T\to\gamma V}(0))^2 +
z^\perp_{33}(M^2_T,M^2_V,0)\ (F^{(3)}_{T\to\gamma V}(0))^2 \right ]\ .
\eea
 The same block of form factors determines the partial width
for $V\to\gamma T$:
\bea
m_V \Gamma_{V\to\gamma T}&=&
e^2\int d\Phi_2(p;,q,p') \frac 13 \sum \limits_{\mu\nu,\alpha\beta}
|A_{\mu\nu,\alpha\beta}|^2
\\ \nn
&=&\frac{\alpha}{12} \frac{m_V^2-m_T^2}{m_V^2}
\left[z^\perp_{11}(M^2_T,M^2_V,0)\ (F^{(1)}_{T\to\gamma V}(0))^2 +
\right .
\\ \nn
&& + \left .
z^\perp_{22}(M^2_T,M^2_V,0)\ (F^{(2)}_{T\to\gamma V}(0))^2 +
z^\perp_{33}(M^2_T,M^2_V,0)\ (F^{(3)}_{T\to\gamma V}(0))^2 \right ]\ .
\nn \eea
Let us emphasize that the factors $z^\perp_{aa}(M^2_T,M^2_V,0)$ are
symmetrical with respect to $T \leftrightarrow V$ permutation:
$z^\perp_{aa}(M^2_T,M^2_V,0)= $ $z^\perp_{aa}(M^2_V,M^2_T,0)$.

\subsection{Transition $A\to\gamma V$}

For the reaction $1^{++}$-$meson\to\gamma 1^{--}$-$meson$ (or
$A\to\gamma V$), one can write
 three partial states: the $S$-wave state
with the total spin ${\bf S}={\bf s}_\gamma+{\bf s}_V=1$ and
 two $D$-wave
states with ${\bf S}={\bf s}_\gamma+{\bf s}_V=1$ .
Generally, we have three spin structures, but only two of
them survive in the case of transverse polarised photon (below, as before,
$p$ is the momentum of the decaying particle and $q$ is that of the
outgoing photon):
\bea
\label{AgV-1}
S_{\mu,\alpha\beta}^{(1)}(p,q)&=&
g_{\alpha\alpha'}^{\perp\perp}g_{\beta\beta'}^{\perp V}\,
\varepsilon_{\mu\alpha'\beta' p}\ ,
\\ \nn
S_{\mu,\alpha\beta}^{(2)}(p,q)&=&
-\frac 1{q_\perp^2}q^{\perp}_{ \beta'}g_{\mu\mu'}^{\perp}
g_{\alpha\alpha'}^{\perp\perp}g_{\beta\beta'}^{\perp V}\,
\varepsilon_{\mu'\alpha' q^\perp p}\ =\
-\frac 1{q_\perp^2}q^{\perp}_{ \beta'}g_{\beta\beta'}^{\perp V}\,
\varepsilon_{\mu\alpha q^\perp p}\ ,
\\ \nn
S_{\mu,\alpha\beta}^{(3)}(p,q)&=&
-\frac 1{q_\perp^2}q^{\perp}_{ \alpha'}g_{\mu\mu'}^{\perp}
g_{\alpha\alpha'}^{\perp\perp}
g_{\beta\beta'}^{\perp V}\;\varepsilon_{\mu'\beta' q^\perp p}=0\ .
\eea
Here, as previously, we use the abridged form
$\varepsilon_{\mu\alpha\beta\xi}p_{\xi}\equiv
\varepsilon_{\mu\alpha\beta p}$ .
The vanishing of $S_{\mu,\alpha\beta}^{(3)}$ is due
to the equality $\;q^{\perp}_{\xi}g_{\alpha\xi}^{\perp\perp}=0$.

\subsubsection{Spin operators and decay amplitude}

The operators $S_{\mu,\alpha\beta}^{(a)}(p,q)$ should be
 orthogonalised:
\bea
\label{AgV-3}
S^{(\perp 1)}_{\mu,\alpha\beta}(p,q)&\equiv
&S^{(1)}_{\mu,\alpha\beta}(p,q)\ , \\
\nn S^{(\perp 2)}_{\mu,\alpha\beta}(p,q)&=
&S^{(2)}_{\mu;\alpha\beta}(p,q) -S^{(\perp 1)}_{\mu,\alpha\beta}(p,q)
 \frac{\left(S^{(\perp 1)}_{\mu',\alpha'\beta'}(p,q)
S^{(2)}_{\mu',\alpha'\beta'}(p,q)\right)}
{\left(S^{(\perp 1)}_{\mu',\alpha'\beta'}(p,q)
S^{(\perp 1)}_{\mu',\alpha'\beta'}(p,q)\right)}
\ .
\end{eqnarray}
We determine the convolutions
\be
\label{AgV-3a}
S^{(\perp a)}_{\mu,\alpha\beta}(p,q)
S^{(\perp b)}_{\mu,\alpha\beta}(p,q)
\equiv z_{ab}^\perp (M^2_A,M^2_V,q^2)\ .
\ee
 At
$q^2=0$ (see Appendix 3 for the details), they are equal to:
\bea
\label{AgV-3b}
z_{11}^\perp (M^2_A,M^2_V,0)&=&\frac
{-(M^4_A+6M^2_AM^2_V+M^4_V)}{2M^2_V}\ ,
\\ \nn
z^\perp_{22}(M^2_A,M^2_V,0)&=&-\frac{2M^2_A(M^2_A+M^2_V)^2}
{(M^4_A+6M^2_AM^2_V+M^4_V)} \ .
\nn
\eea
The transition amplitude $A\to \gamma V$ reads:
\be
\label{AgV-5}
A^{(A\to\gamma V(L'))}_{\mu,\alpha\beta}\ =\
\sum_{i=1,2} S^{(\perp i)}_{\mu,\alpha\beta}(p,q)
F^{(i)}_{A\to\gamma V(L)}(0)\ ,
\ee
being determined by two
form factors  $F^{(i)}_{A\to\gamma V}(0)$ ($i=1,2$).

\subsubsection{Calculation of the quark triangle diagram of Fig. 1
for the  emission of the real photon}

The diagram of Fig. 1b for the processes $A\to\gamma V(L)$
($L=0,2$) is determined by the following traces:
\bea
\label{AgV-6}
Sp_{\mu,\alpha\beta}^{(A\to\gamma V(0))}&= &
-Sp\left[\hat G^{(1,0,1)}_\beta(\hat{k'_1}+m)\gamma_\alpha^{\perp\gamma}
(\hat{k_1}+m)T_\mu(k)(-\hat{k_2}+m)\right],
\\ \nn
Sp_{\mu,\alpha\beta}^{(A\to\gamma V(2))}&= &
-Sp\left[\hat G^{(1,2,1)}_\beta(\hat{k'_1}+m)\gamma_\alpha^{\perp\gamma}
(\hat{k_1}+m)T_\mu(k)(-\hat{k_2}+m)\right]\ ,
\eea
where the transition $A\to Q\bar Q$ is equal to:
\bea
\label{AgV-7}
T_\mu(k)&=&\sqrt{\frac{2}{3s}}i\,\varepsilon_{\mu k \gamma P}\ .
\eea
Recall that $k=(k_1-k_2)/2$ and $P=k_1+k_2$.

To calculate the invariant form factor $F_{A\to\gamma V(L)}(q^2)$, we
should expand (\ref{AgV-6}) in a series with respect to the spin
operators $S^{(\perp i)}_{\mu,\alpha\beta}(P, \tilde q)$ where $\tilde
q=P-P'$:
\bea \label{AgV-9}
Sp^{(A\to\gamma V(L))}_{\mu,\alpha\beta} &=& \sum_{i=1,2}
S^{(i\perp)}_{\mu,\alpha\beta}(P,\tilde q)
S^{(i)}_{A\to\gamma V(L)}(s,s',q^2)\ ,
\eea
that gives us
\bea
\label{AgV-10}
S^{(i)}_{A\to\gamma V(L)}(s,s',q^2)&=&
\frac{\left (Sp^{(A\to\gamma V(L))}_{\mu,\alpha\beta}
        S^{(i\perp)}_{\mu,\alpha\beta}(P,\tilde q)\right )}
{\left (S^{(i\perp)}_{\mu,\alpha\beta}(P,\tilde q)
        S^{(i\perp)}_{\mu,\alpha\beta}(P,\tilde q)\right )}\ ,
\eea
where $i=1$ or $2$.

The transition form factor is determined by the standard formula:
\bea
\label{AgV-11}
F^{(i)}_{A\to\gamma V(L)}(q^2)&=&Z_{A\to\gamma V}
\int \limits_{4m^2}^\infty \frac{dsds'}{16\pi^2}
\psi_{A}(s)\psi_{V(L)}(s')\times
\\ \nn
&&\times\frac{\theta(-ss'q^2-m^2\lambda(s,s',q^2))}
{\sqrt{\lambda(s,s',q^2)}} S^{(i)}_{A\to\gamma V(L)}(s,s',q^2) ,
\eea
In the limit $q^2\to 0$,
performing calculations as in the cases considered above, we
obtain:
\bea
\label{AgV-12}
F^{(i)}_{A\to\gamma V(L)}&=&Z_{A\to\gamma V}
\int \limits_{4m^2}^\infty \frac{ds}{16\pi^2}(s)
\psi_{A}(s)\psi_{V(L)}(s)J^{(i)}_{A\to\gamma V(L)}(s)\ ,
\\ \nn
J^{(1)}_{A\to\gamma V(0)}(s)&=&-\sqrt{\frac32} I_{A\to\gamma V}(s),
\\ \nn
J^{(2)}_{A\to\gamma V(2)}(s)&=& \frac{\sqrt 3}{8}(4m^2-s)
I_{A\to\gamma V}(s),
\eea
where
\be
\label{AgV-13}
I_{A\to\gamma V}(s)=
\sqrt{s}\left(2m^2\ln\frac{\sqrt{ s}+\sqrt {s-4m^2}}
             {\sqrt{ s}-\sqrt {s-4m^2}}-\sqrt {s(s-4m^2)}\right).
\ee
The total form factor is equal to:
\be
F_{A\to\gamma V}(0)\ =\ F_{A\to\gamma V(0)}(0)+F_{A\to\gamma V(2)}(0)
\ee

\subsubsection{Normalisation condition and partial widths}

The normalisation condition for the $1^{++}$ meson wave function reads:
\bea
\label{AgV-14}
1= \frac12\int \limits_{4m^2}^\infty \frac {ds}{16\pi^2} \,
\psi_{A}^2(s)\; s(s-4m^2)\rhosq.
\eea
The partial width of the decay $A\to\gamma V$ is equal to
\bea
\label{AgV-15}
m_{A} \Gamma_{A\to\gamma V}
&=& e^2\int d\Phi_2(p;,q,p') \frac 13
\sum \limits_{\mu,\alpha\beta} |A_{\mu,\alpha\beta}|^2
\\ \nonumber
&=&\frac {\alpha} {12}\frac{m_{A}^2-m_{V}^2}{m_{A}^2}\;
\left[
z^\perp_{11}(M^2_A,M^2_V,0)\left(F^{(1)}(0)\right)^2+
z^\perp_{22}(M^2_A,M^2_V,0)\;\left(F^{(2)}(0)\right)^2\right]\ .
\eea
For the
partial width of the decay $V\to\gamma A$, one has:
\bea
m_{V} \Gamma_{V\to\gamma A}
&=&\frac {\alpha} {12}\frac{m_{V}^2-m_{A}^2}{m_{V}^2}\;
\left[
z^\perp_{11}(M^2_V,M^2_A,0)\left(F^{(1)}(0)\right)^2+
\right .
\\ \nn
&+& \left .
z^\perp_{22}(M^2_V,M^2_A,0)\;\left(F^{(2)}(0)\right)^2\right] .
\eea
Let us emphasise that  $z^\perp_{aa}(M^2_V,M^2_A,0)\neq
z^\perp_{aa}(M^2_A,M^2_V,0)$.

\section{Conclusion}

The considerations of the tensor meson decays (Section 3) and pseudovector
meson decay (Section 4), being in fact general cases, can be easily expanded
 for any mesons.

Actually, the tensor meson decay is a pattern for an amplitude,
where the parity of initial meson coincides with the parity of final
state. For this case, we construct the spin scalars from the
polarisation and angular momentum functions
$X^{(L)}_{\mu_1\cdots\mu_L}(k^\perp)$,
see Eq. (\ref{TgV-2}) for tensor meson. With a completness condition
for the vector and tensor polarisations, we construct
gauge invariant spin operators ((\ref{TgV-3}) for the tensor mesons.
The orthogonalisation of these operators for the case of the real photon
emission allows us to single out the operators with nonzero norm, of
the type of Eq. (\ref{TgV-4})), and nilpotent operators.
The operators of the first kind are used in the expansion of the
amplitude in a series  in respect to external operators (Eq.
(\ref{TgV-7})), as well as for  the quark
triangle diagram (Eqs. (\ref{dsi-8}) and (\ref{dsi-9})). The spectral
integrals are written for the invariant form factors, which are the
coefficients in front of the orthogonalised operators.

The decay of axial meson provides us with an example, where the parity
changes sign from initial to final state. In this case, there is only
one difference: in the first step, we construct the pseudoscalars from the
particle polarisations and  angular
momentum function $X^{(L)}_{\mu_1\cdots\mu_L}(k^\perp)$.
Further consideration is carried out following the  scheme common to
that of tensor mesons.

\section*{Acknowledgments}
We thank Y.I. Azimov, L.G. Dakhno and D.I. Melikhov for useful
discussion of the spectral integral representation of amplitudes.
This work was supported by the Russian Foundation for Basic Research,
project 04-02-17091 .

\section*{Appendix 1: Convolutions of spin operators\\ in
the decay $T\to\gamma V$}
In the calculation of the spin factors we need the convolutions
\be
S_{\mu\nu,\alpha\beta}^{(a)}(P',\tilde q)
S_{\mu\nu,\alpha\beta}^{(b)}(P',\tilde q)
\equiv z_{ab}(s,s',q^2)\ .
\ee
They are as follows:
\bea
z_{11}(s,s',q^2)&=&\frac{3s^2-6s q^2+34ss'+3q^4-6q^2 s'+3s'^2}{12ss'}\ ,
\\ \nn
z_{22}(s,s',q^2)&=&3\ ,
\\ \nn
z_{33}(s,s',q^2)&=&\frac{3s^2-6s q^2+13ss'+3q^4-6q^2 s'+3s'^2}{12ss'}\ ,
\nn \\
z_{44}(s,s',q^2)&=&\frac{3s^2-6sq^2+34ss'+3q^4-6q^2s'+3s'^2}{48ss'}\ ,
\nn \\
z_{12}(s,s',q^2)&=& 1\ ,
\nn \\
z_{13}(s,s',q^2)&=&-\frac{3s^2-6s q^2-8ss'+3q^4-6q^2s'+3s'^2}{12ss'}\ ,
\nn \\
z_{14}(s,s',q^2)&=&\frac{3s^2-6sq^2+34ss'+3q^4-6q^2s'+3s'^2}{24ss'}\ ,
\nn \\
z_{23}(s,s',q^2)&=&\frac 12\ ,
\nn \\
z_{24}(s,s',q^2)&=&\frac 12\ ,
\nn \\
z_{34}(s,s',q^2)&=&\frac{-3s^2+6sq^2+8ss'-3q^4+6q^2s'-3s'^2}{24ss'}\ ,
\nn \\
z_{55}(s,s',q^2)&=&\frac{2s^2-4sq^2+11ss'+2q^4-4q^2s'+2s'^2}{8ss'}\ ,
\nn \\
z_{15}(s,s',q^2)&=&\frac{-s^2+2sq^2+2ss'-q^4+2q^2s'-s'^2}{4ss'}\ ,
\nn \\
z_{25}(s,s',q^2)&=&\frac 32\ ,
\nn \\
z_{35}(s,s',q^2)&=&\frac{s^2-2sq^2+4ss'+q^4-2q^2s'+s'^2}{4ss'}\ ,
\nn \\
z_{45}(s,s',q^2)&=&-\frac{s^2-2sq^2-2ss'+q^4-2q^2s'+s'^2}{8ss'}\ .
\eea
For the orthogonal operators, we have:
\be
S_{\mu\nu,\alpha\beta}^{(\perp a)}(P',\tilde q)
S_{\mu\nu,\alpha\beta}^{(\perp b)}(P',\tilde q)\equiv z^\perp_{ab}
(s,s',q^2) \
\ee
with
\bea
z^\perp_{11}(s,s',q^2)&=&\frac{3s^2-6s q^2+34ss'+3q^4-6q^2 s'+3s'^2}{12ss'}\ ,
\\ \nn
z^\perp_{22}(s,s',q^2)&=&9\frac{s^2-2sq^2+10ss'+q^4-2q^2s'+s'^2}
{3s^2-6sq^2+34ss'+3q^4-6q^2s'+3s'^2}
\\ \nn
z^\perp_{33}(s,s',q^2)&=&\frac92
\frac{(s-q^2+s')^2}{s^2-2s q^2+10ss'+q^4-2q^2s'+s'^2}\ .
\eea
Spin factors given by $\Sigma,z$ variables in the limit
$Q\to 0$ read:
\bea
S^{(1)}_{T(1)\to\gamma V(0)}(\Sigma,z,Q\to  0)&=&-\frac{3\sqrt 2}{5}
\frac{(4 m^2 \Sigma+m^2 z^2-\Sigma^2) (8 m^2+3 \Sigma)}
  {4 \Sigma+z^2},  \\
\nn
S^{(2)}_{T(1)\to\gamma V(0)}(\Sigma,z,Q\to  0)&=& -\frac{8\sqrt 2}{3}
\frac{ (4 m^2 \Sigma+m^2 z^2-\Sigma^2+\Sigma z^2) (4 m^2 \Sigma+m^2 z^2-\Sigma^2)}
 {(4 \Sigma+z^2)^2},
 \\ \nn
S^{(1)}_{T(1)\to\gamma V(2)}(\Sigma,z,Q\to  0)&=& -\frac{3}{20}
\frac{ (4 m^2 \Sigma+m^2 z^2-\Sigma^2) (16 m^2-3 \Sigma) (4 m^2-\Sigma)}
 {4 \Sigma+z^2},
 \\ \nn
S^{(2)}_{T(1)\to\gamma V(2)}(\Sigma,z,Q\to  0)&=& -\frac{2}{3}
\frac{(4 m^2 \Sigma+m^2 z^2-\Sigma^2+\Sigma z^2)
 (4 m^2 \Sigma+m^2 z^2-\Sigma^2) (8 m^2+\Sigma)}
{(4 \Sigma+z^2)^2},
\\ \nn
S^{(1)}_{T(3)\to\gamma V(0)}(\Sigma,z,Q\to 0)&=& -\frac{3\sqrt 2}{10}
\frac{(4 m^2 \Sigma+m^2 z^2-\Sigma^2) (4 m^2-\Sigma)^2}
{4 \Sigma+z^2},
\\ \nn
S^{(2)}_{T(3)\to\gamma V(0)}(\Sigma,z,Q\to 0)&=& -\frac{4\sqrt 2}{9}
\frac{(4 m^2 \Sigma+m^2 z^2-\Sigma^2+\Sigma z^2)
    (4 m^2 \Sigma+m^2 z^2-\Sigma^2) (6 m^2+\Sigma)}
{(4 \Sigma+z^2)^2},
\\ \nn
S^{(1)}_{T(3)\to\gamma V(2)}(\Sigma,z,Q\to 0)&=& -\frac{3}{40}
\frac{(4 m^2 \Sigma+m^2 z^2-\Sigma^2) (8 m^2+\Sigma) (4 m^2-\Sigma)^2}
{4 \Sigma+z^2},
\\ \nn
S^{(2)}_{T(3)\to\gamma V(2)}(\Sigma,z,Q\to 0)&=&
\frac{ (4 m^2 \Sigma+m^2 z^2-\Sigma^2+\Sigma z^2)
 (4 m^2 \Sigma+m^2 z^2-\Sigma^2) (12 m^2-\Sigma) (\Sigma -4 m^2)}
 {9(4 \Sigma+z^2)^2} ,
\\ \nn
S^{(3)}_{T(L)\to\gamma V(L')}(\Sigma,z,Q\to 0)&=&
 -\frac{2}{3}S^{(2)}_{T(L)\to\gamma V(L')}(\Sigma,z,Q\to 0)\ .
\eea

\section*{Appendix 2: Convolutions of spin operators\\ in
the decay $A\to\gamma V$}

We denote the convolutions of the spin operators at $q^2\ne 0$ as
follows:
\be
\label{AgV-2}
S_{\mu;\alpha\beta}^{(a)}(P',\tilde
q)S_{\mu;\alpha\beta}^{(b)} (P',\tilde q)\equiv z_{ab}(s,s',q^2)\ ,
\ee
where
\bea
z_{11}(s,s',q^2)&=&\frac
{-(s^2+6ss'+s'^2)+q^2(2s+2s'-q^2)}{2s'}\ ,\nn
\\
z_{22}(s,s',q^2)&=&-\frac{(s+s'-q^2)^2}{2s'}\ ,
\nn \\
z_{12}(s,s',q^2)&=&\frac{(s+s'-q^2)^2}{2s'}\ .
\eea
For the orthogonalised operators,
\be
\label{AgV-4}
S_{\mu;\alpha\beta}^{(\perp a)}(P',\tilde q)
S_{\mu;\alpha\beta}^{(\perp b)}(P',\tilde q)\equiv z^\perp_{ab}(s,s',q^2)\ ,
\ee
one obtains:
\bea
z_{11}^\perp (s,s',q^2)&=&\frac
{-(s^2+6ss'+s'^2)+q^2(2s+2s'-q^2)}{2s'}\ ,\nn
\\
z^\perp_{22}(s,s',q^2)&=&-\frac{2s(s+s'-q^2)^2}
{(s^2+6ss'+s'^2)-q^2(2s+2s'-q^2)} \ .
\eea
For the calculation of the spin factors in the double dispersion integral,
we use variables
$s=\Sigma+\frac 12 zQ$,
$s'=\Sigma-\frac 12 zQ$, and $q^2=-Q^2$. Using these variables, in the
limit  $Q\to 0$, we have for spin factors:
\bea
S^{(1)}_{A\to\gamma V(0)}(\Sigma,z,Q\to 0)&=&\frac{5\sqrt 2}{\sqrt 3}
\Sigma\frac{m^2 z^2+4  m^2 \Sigma-\Sigma^2}
{\sqrt{\Sigma}(z^2+4\Sigma)}\ ,
\\ \nn
S^{(2)}_{A\to\gamma V(0)}(\Sigma,z,Q\to 0)&\to&\; Qz
\left [-\frac{5}{\sqrt 6}\frac{(m^2z^2+4m^2\Sigma-3\Sigma^2)}
{\sqrt{\Sigma}(z^2+4\Sigma)}\right ]\ \to\ 0\ ,
\\ \nn
S^{(1)}_{A\to\gamma V(2)}(\Sigma,z,Q\to 0)&=& \frac{\sqrt 2}{\sqrt 3}
\Sigma^2\frac{(4m^2\Sigma+m^2z^2-\Sigma^2)(\Sigma-4m^2) }
{\sqrt{\Sigma}(z^2+4\Sigma)}\ ,
\\ \nn
S^{(2)}_{A\to\gamma V(2)}(\Sigma,z,Q\to 0)&\to&\; Qz  \left [
-\frac{1}{2\sqrt 6}
\frac{(32m^4\Sigma+8m^4z^2+4m^2\Sigma^2+7m^2\Sigma z^2-3\Sigma^3)}
{\sqrt{\Sigma}(z^2+4\Sigma)}\right ] \to\ 0\ .
\eea

\end{document}